\documentclass[aps,preprint,superscriptaddress]{revtex4-1}
\usepackage[margin=1in]{geometry}
\usepackage[utf8x]{inputenc}
\usepackage{amssymb}
\usepackage{latexsym}
\usepackage{amsfonts}
\usepackage{graphicx}
\usepackage{epsfig}
\usepackage{amsmath}
\usepackage[retainorgcmds]{IEEEtrantools} 
\usepackage{float}
\usepackage{verbatim} 
\usepackage{subfigure}
\usepackage{color} 
\usepackage{ulem}
\usepackage{mathtools}
\usepackage{epstopdf}

\graphicspath{{./figs/}}

\def\kcore{$k$-core }

\def\er{Erd\H{o}s-R\'enyi }

\begin{document}

\title{Generalized model for \kcore percolation and interdependent networks} 

\author{Nagendra K. Panduranga}
\affiliation{Center for Polymer Studies and Department of Physics,
Boston University, Boston, Massachusetts 02215 USA}

\author{Jianxi Gao}
\affiliation{Center for Complex Network Research and Department of
Physics, Northeastern University, Boston, Massachusetts 02115, USA}

\author{Xin Yuan}
\affiliation{Center for Polymer Studies and Department of Physics,
Boston University, Boston, Massachusetts 02215 USA}

\author{H. Eugene Stanley}
\affiliation{Center for Polymer Studies and Department of Physics,
Boston University, Boston, Massachusetts 02215 USA}

\author{Shlomo Havlin}
\affiliation{Department of Physics, Bar-Ilan University, Ramat-Gan 52900,
Israel}

\begin{abstract}

\noindent
Cascading failures in complex systems have been studied extensively using two different models: \kcore percolation and interdependent networks. We combine the two models into a general model, solve it analytically and validate our theoretical results through extensive simulations. We also study the complete phase diagram of the percolation transition as we tune the average local \kcore threshold and the
coupling between networks. We find that the phase diagram of the combined processes  
is very rich and includes novel features that do not appear in the models studying each of the processes separately. For example, the phase diagram consists of 
first and second-order transition regions separated by two tricritical
lines that merge together and enclose a novel two-stage transition
region. In the two-stage transition, the size of the giant component
undergoes a first-order jump  at a certain occupation probability followed by a continuous second-order
transition at a lower  occupation probability. Furthermore, at certain fixed interdependencies, the percolation transition changes from first-order $\rightarrow$ second-order $\rightarrow$ two-stage $\rightarrow$ first-order as the \kcore threshold is increased. The analytic
equations describing the phase boundaries of the two-stage transition
region are set up and the critical exponents for each type of transition
are derived analytically.

\end{abstract}

\maketitle

Understanding cascading failures is one of the central questions in the
study of complex systems \cite{taming_complex}. In complex systems, such as power grids \cite{Scala},
financial networks \cite{krugman_econ_casc}, and social systems \cite{vespi_social}, even a
small perturbation can cause sudden cascading failures. In particular,
two models for cascading failures with two different mechanisms were
studied extensively and separately, \kcore percolation \cite{wattspnas,
Gleeson_pnas} and interdependency between networks \cite{SergeyNature, Parshani_prl, vespi_intdep, 
ieee_dep}.

In single networks, \kcore is defined as a maximal set of nodes that
have at least $k$ neighbors within the set. The algorithm to find
$k$-cores is a local process consisting of repeated removal of nodes
having fewer than $k$ neighbors until every node meets this
criterion. This process of pruning nodes can be mapped to one of the
causes for cascading failures \cite{wattspnas,Gleeson_pnas}. For
example, after some initial damage to a power grid network, nodes with
fewer than a certain number of neighbors can fail due to electric power
overload \cite{Raisa2}. This scenario corresponds to \kcore
percolation. The threshold $k$ can be node-dependent, which is often
referred to as heterogeneous \kcore percolation. Both homogeneous and
heterogeneous cases have been extensively studied in single networks
\cite{prl_dor,htrgnskcore1,tricrit,crit_hkc}. 

Another salient feature of real-world systems that causes cascading
failures is interdependency. For example, power network and
communication network depend on each other to function and regulate, so
failure in one network or both networks leads to cascading failures in
one or both systems. Cascading failures have been studied extensively as
percolation in interdependent networks \cite{SergeyNature, Parshani_prl,
Zhou_SF, JianxiNature, Boccaletti, son2012percolation}. Increase in either interdependency or $k$-core threshold increases the instability in networks. The models, studying these processes separately, demonstrate this with percolation transition changing from second-order $\rightarrow$ first-order as the parameters are increased \cite{Parshani_prl, tricrit}. 
  
Here we combine both processes (\kcore percolation and interdependency) into a single general model to study the combined effects. We demonstrate that the results of the combination are very rich and include novel features that do not appear in the models that study each
process separately. Furthermore, some results are counterintuitive to the results from studying the processes separately. For example, at certain fixed interdependencies, the percolation transition changes from first-order $\rightarrow$ second-order $\rightarrow$ two-stage $\rightarrow$ first-order as the \kcore threshold is increased.

\begin{figure}
\includegraphics[width=0.8\columnwidth]{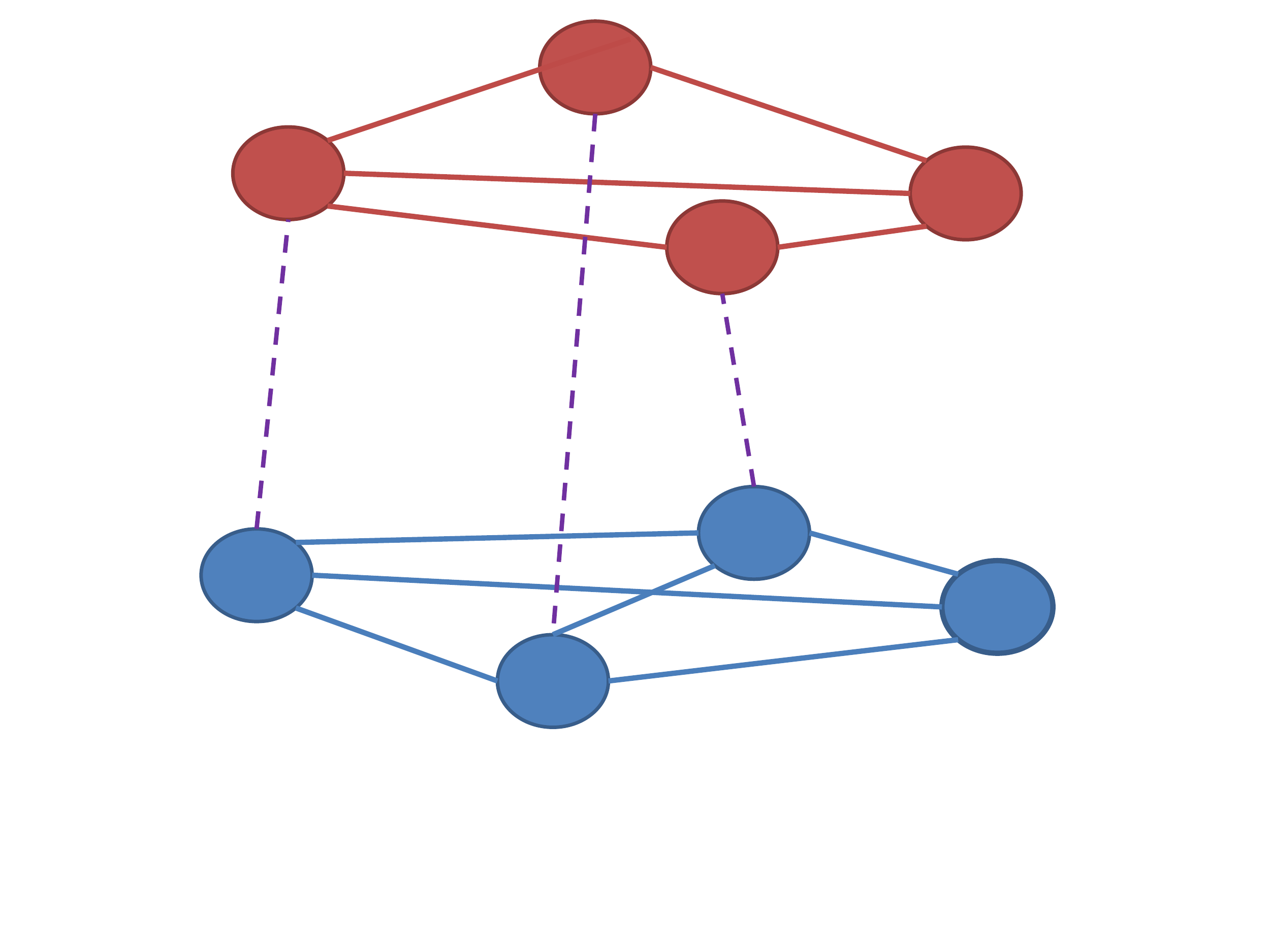}
\caption{Demonstration of an interdependent network with coupling $q =
  0.75$ with dependency links shown as dashed lines. The $2$-core and
  $3$-core are the highest possible \kcore in the top and bottom layers
  respectively, while still preserving all the dependency
  links.} \label{fig:schem}
\end{figure}

Consider a system composed of two interdependent uncorrelated random
networks A and B with both having the same arbitrary degree distribution
$P(i)$. The coupling $q$ between networks is defined as the fraction of
nodes in network A depending on nodes in network B and vice versa
(Fig.~\ref{fig:schem}). The \kcore percolation process is initiated by
removing a fraction $1-p_0$ of randomly chosen nodes, along with all
their edges, from both networks. In \kcore percolation, nodes in the
first network with fewer than $k_a$ neighbors are pruned (the local
threshold of each node may differ), along with all the nodes in the second
network that are dependent on them. The \kcore percolation process is
repeated in the second network, and this reduces the number of neighbors
of nodes in the first network to fewer than $k_a$. This cascade process
is continued in both networks until a steady state is reached. The cascades 
in both networks are bigger during \kcore peroclaiton than during regular percolation 
due to pruning process. Here we consider the case of heterogeneous \kcore percolation in which a fraction $r$ of
randomly chosen nodes in each network is assigned a local threshold
$k_a+1$ and the remaining fraction $1-r$ nodes are assigned a threshold
$k_a$. This makes the average local threshold per site, identical for
both networks, to be $k = (1-r)k_a + r(k_a+1)$, which allows us to study
the \kcore percolation continuously from $k_a$-core to ($k_a+1$)-core by
changing the fraction $r$. Note that the \kcore percolation properties
depend on the distribution of local thresholds $k_a$, and not on the
average threshold per site as found in single networks
\cite{ph_transt_kor, crit_hkc}.

At the steady state of the cascade process the network becomes
fragmented into clusters of various sizes. Only the largest cluster (the
``giant component'') is considered functional in this study and is the
quantity of interest. The fraction of nodes $\phi^{\prime}_{\infty}$
remaining in the steady state is identical in both networks as the
entire process is symmetrical for both networks and can be calculated
using the formalism developed by Parshani \textit{et al} \cite{Parshani_prl},

\begin{equation}
  \phi^{\prime}_{\infty}\equiv
  p_0[1-q(1-p_0M_{k_a,r}(\phi^{\prime}_{\infty}))], \label{eqn:phi-psi} 
\end{equation}
where $M_{k_a,r}(p)$ is the fraction of nodes belonging to the giant
component in a single network with a fraction $p$ of nodes occupied. The
size of the giant component in the coupled networks at the steady state
$\phi_{\infty}$ is
\begin{equation}\label{eqn:phi_inf}
  \phi_{\infty} =  \phi^{\prime}_{\infty}M_{k_a,r}(\phi^{\prime}_{\infty}).
\end{equation}

The \kcore formalism for single networks \cite{htrgnskcore1}, based on
local tree-like structure, gives the size of the giant component,
\begin{IEEEeqnarray}{rCl}
	M_{k_a,r}(p)  &=&   (1-r)\sum_{j=k_a}^{\infty}
P(j) \Phi_{j}^{k_a}(X(p),Z(p)) +\nonumber\\
	&&  r\sum_{j=k_a+1}^{\infty} P(j)
\Phi_{j}^{k_a+1}(X(p),Z(p)), \label{eq:mass-k}
\end{IEEEeqnarray}
where
\begin{equation}
	\Phi_{j}^{k}(X,Z) = \sum_{l=k}^{j} {j \choose l} (1-X)^{j-l}
\sum_{m=1}^l {l \choose m} Z^m (X-Z)^{l-m} \nonumber.
\end{equation}

Here $Z$ and $X$ are the probabilities that, starting from any random
link and node, respectively, the giant component will be reached. These
are calculated using the self-consistent equations
\begin{IEEEeqnarray}{rCl}
	\frac{X}{f_{k_a,r}(X,X)} = \frac{Z}{f_{k_a,r}(X,Z)} =
        p, \label{eqn:XZ} 
\end{IEEEeqnarray}
where
\begin{IEEEeqnarray}{rCl}
	 f_{k_a,r}(X,Z) &=& (1-r) \sum_{j=k_a}^{\infty}
         \case{jP(j)}{\langle j 
\rangle} \Phi_{j-1}^{k_a-1}(X,Z) +\nonumber\\
&& r \sum_{j=k_a+1}^{\infty} \case{jP(j)}{\langle j \rangle}
\Phi_{j-1}^{k_a}(X,Z).	\label{eq:fk}
\end{IEEEeqnarray}
The probabilities $X$ and $Z$ are equal when the local thresholds of
\kcore percolation are $k_a\ge2$ \cite{tricrit}.
Equations (\ref{eqn:phi-psi}) and (\ref{eq:mass-k})--(\ref{eq:fk}) can
be further simplified,
\begin{equation}\label{eqn:hkz}
 p_0 = \frac{q-1 + \sqrt{(q-1)^2 + 4 q
\frac{Z M_{k_a,r}(X(Z),Z)}{f_{k_a,r}(Z,Z)}}
}{2 q M_{k_a,r}(X(Z),Z)}\equiv h_{k,q}(Z), 
\end{equation}
which can be used to solve for the probabilities $X$, $Z$ for any
initial percolation probability $p_0$.  The size of the giant component
as a function of $p_0$, found by numerically solving
Eqs.~(\ref{eqn:phi_inf})--(\ref{eq:mass-k}) and (\ref{eqn:hkz}), is in
excellent agreement with simulation results for both \er (see
Fig.~\ref{fig:simvthr}) and scale-free networks (see
Fig.~\ref{fig:simvthr_sf} in the Supplementary Material).

\begin{figure}
 \begin{center}
\mbox{
\subfigure{\includegraphics[scale=0.22]{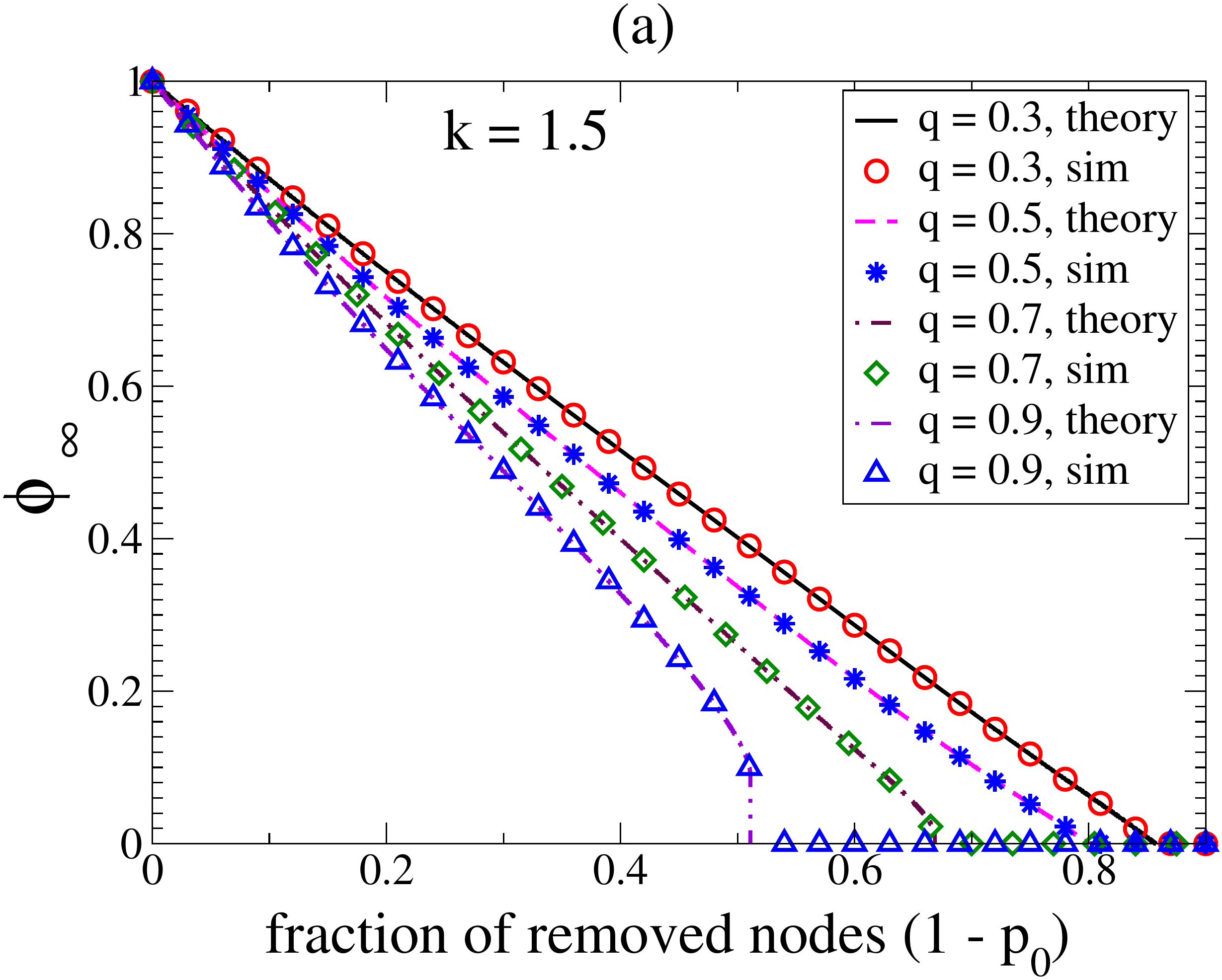}}}
\hspace{.0in} 
\mbox{
\subfigure{\includegraphics[scale=0.22]{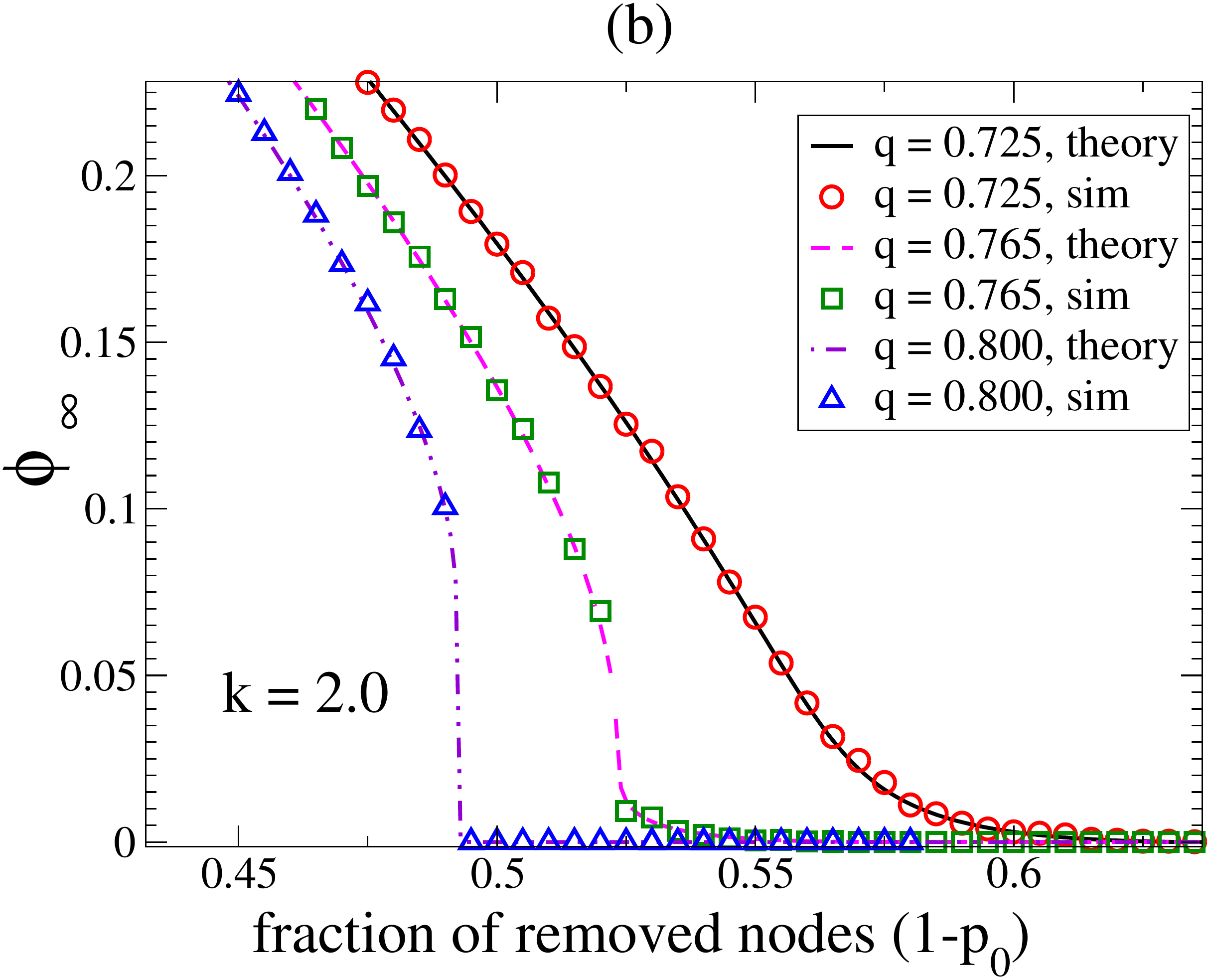}}}
\hspace{.0in} 
\subfigure{\includegraphics[scale=0.22]{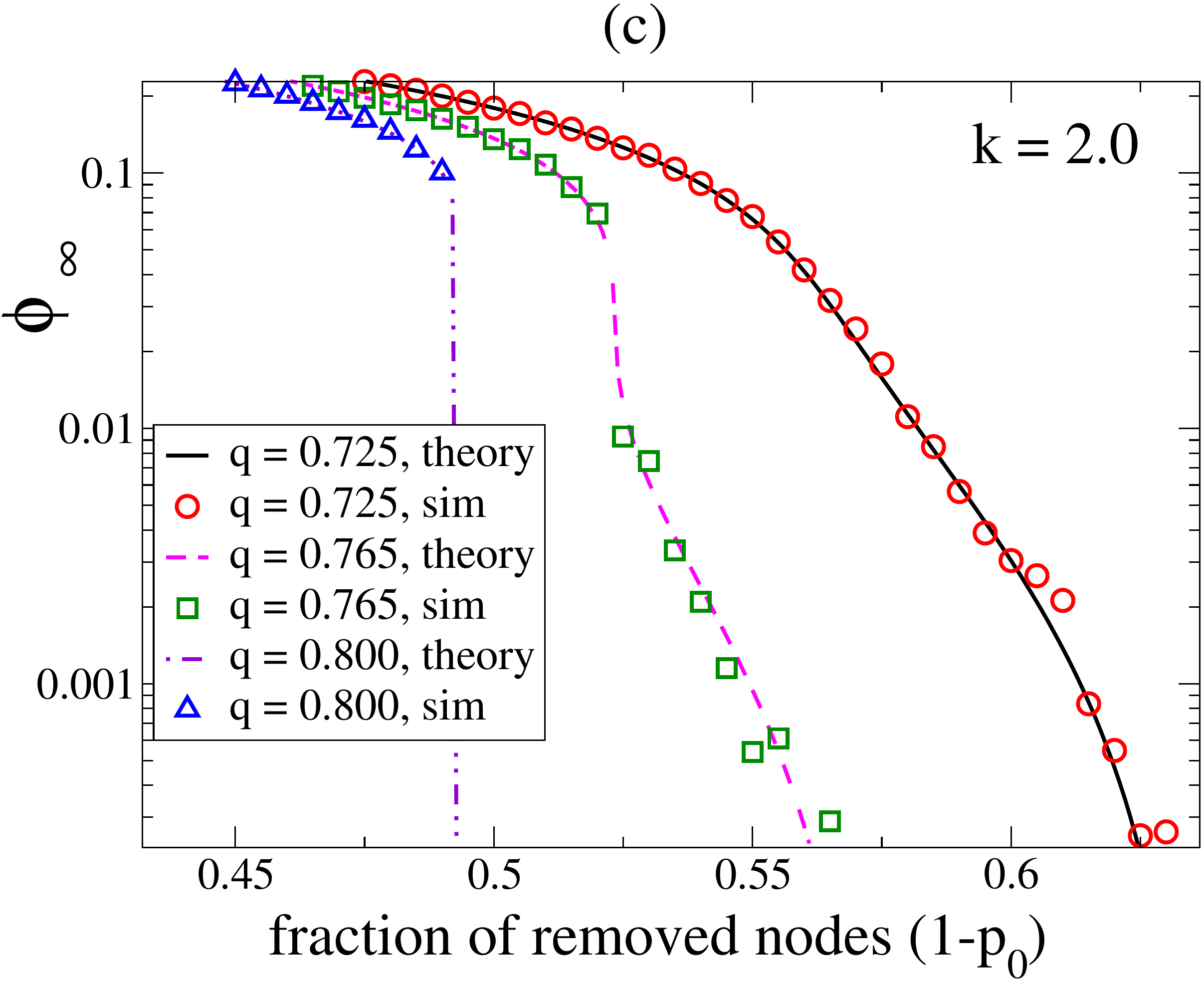}}
\hspace{.0in} 
\subfigure{\includegraphics[scale=0.22]{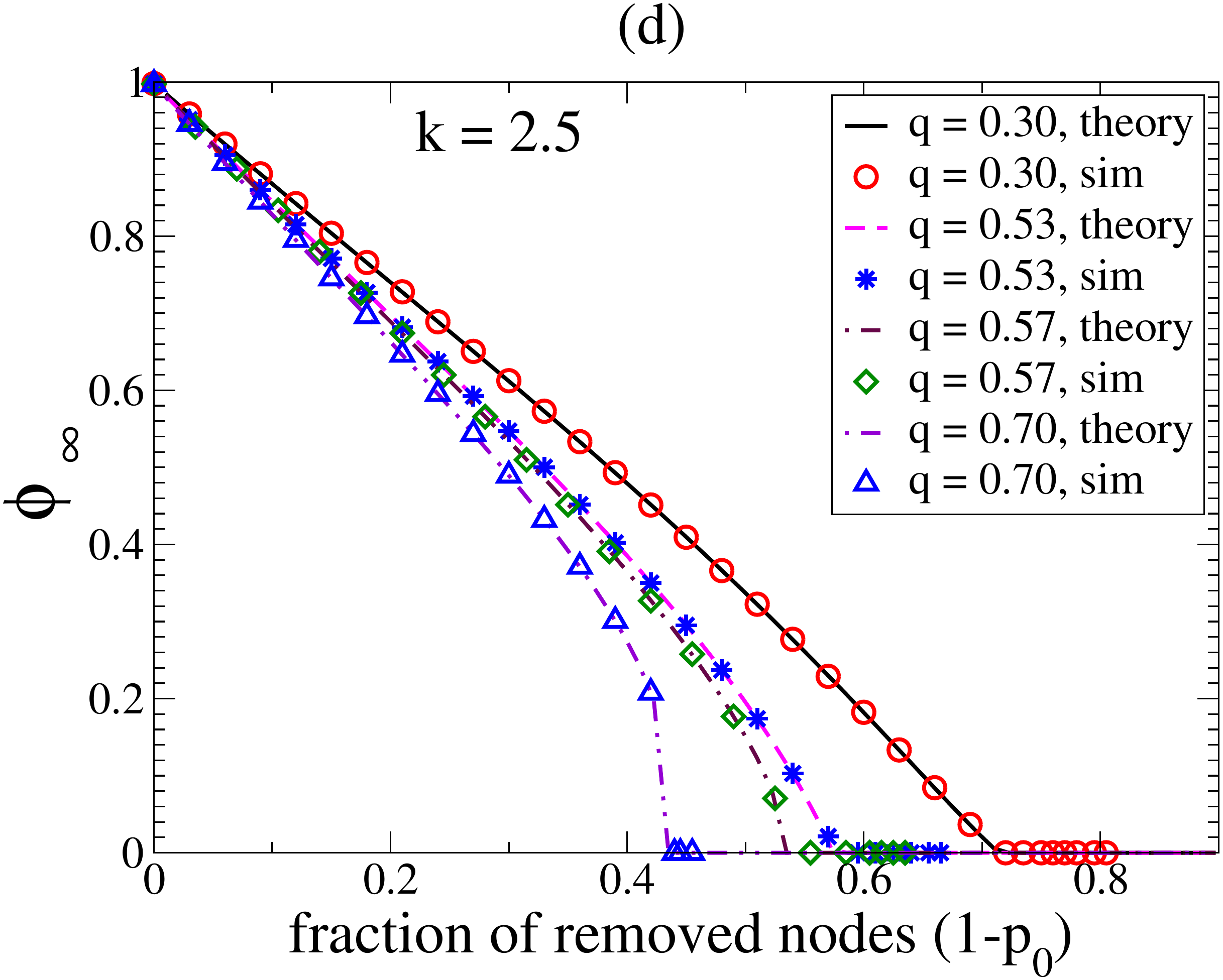}}
\caption{Comparison of theory (lines) and simulation (symbols) for two
  coupled \er networks at fixed average local threshold a) $k = 1.5$ b)
  $k = 2.0$ c) $k = 2.0$ (on a semi-log plot) and d) $k = 2.5$. As the
  coupling $q$ is increased, \kcore percolation transition changes from
  second-order to first-order. For $k = 2.0$, a novel two-stage
  transition is seen at intermediate couplings. Simulation results agree
  well with the theory (more evident in the semi-log plot of panel
  (c)). Plots of comparison for more parameter values are given in the
  Supplementary Material (Fig. \ref{fig:Svpfixedq}).}\label{fig:simvthr}
 \end{center}
\end{figure}

The function $h_{k,q}(Z)$ in Eq.~(\ref{eqn:hkz}) determines the nature
of the phase transition and the critical percolation thresholds $p_c$,
illustrated below in the example of two \er networks.

To demonstrate the richness of the model that combines $k$-core and
interdependency, we focus on two interdependent \er networks. Both
networks have identical degree distributions given by
$P(i)=z_{1}^{i}\exp(-z_{1})/i!$ with the same average degree $z_1$. The
function $f_{k_a,r}$ is given by $f_{1,r}(X,Z) = 1 - e^{-z_{1}Z} $,
$f_{1,r}(X,X) = 1 - re^{-z_{1}X} $ and since $X = Z$ for $k_a \ge 2$,
$f_{2,r}(Z,Z) = 1 - e^{-z_{1}Z} ( 1+rz_{1} Z)$. The functions
$M_{k_a,r}$ are given by $M_{1,r}(X,Z) = 1-e^{-z_1Z}-rz_1Ze^{-z_1X}$ and
$M_{2,r}(Z) = 1-(1-r)\frac{\Gamma(2,z_1Z)}{\Gamma(2)} -
r\frac{\Gamma(3,z_1Z)}{\Gamma(3)}$, where $\Gamma(m,x)$ and $\Gamma(m)$
are incomplete and complete gamma functions, respectively, of order $m$.

\begin{figure}
 \begin{center} 
\includegraphics[scale=0.25]{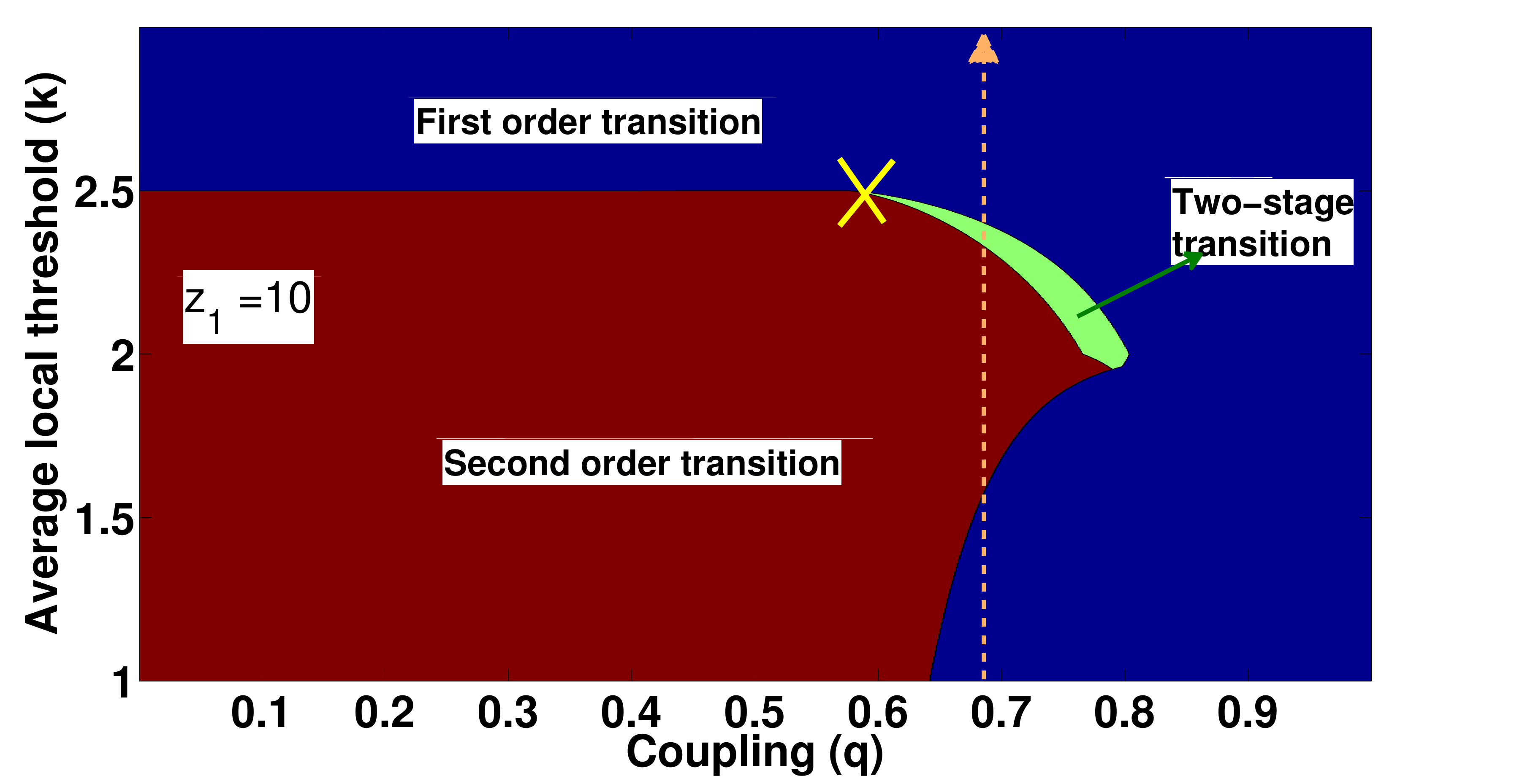} 
\caption{Complete phase diagram for \kcore percolation transition for
  two interdependent \er networks with average degree $z_1 = 10$ . Both
  networks have the same average local threshold per site
  $k=(1-r)k_a+r(k_a+1)$, with fraction $1-r$ of randomly chosen nodes
  having local threshold $k_a$ and the remaining nodes having local
  threshold $k_a+1$. Plots depicting the size of giant component as a
  function of removed nodes for different points in the phase diagram
  are given in Fig. \ref{fig:simvthr} and Fig. \ref{fig:Svpfixedq} in
  the Supplementary Material. Symbol 'X' in the
  phase diagram indicates the coupling $q_{c,2.5}$. The transition
  properties depend on the composition of the $k_a$-cores and not on
  average $k$. For example, the average local threshold of $3$ can be
  achieved by setting half of the nodes with local threshold $2$ and
  remaining nodes with local threshold $4$. \kcore percolation of this
  heterogeneous case is found to be different from that of the
  homogeneous case where all the nodes have the same threshold
  $3$. Phase diagrams for different average degree $z_1$ are shown in
  the Supplementary Material
  (Fig. \ref{fig:phaseplot_diffz_ER}).} \label{fig:phaseplot}
\end{center} 
\end{figure}

The behavior of the function $h_{k,q}(Z)$, Eq.~(\ref{eqn:hkz}), for fixed
values of parameters as a function of $Z$ determines the nature of the
\kcore percolation transition. In general, the function $h_{k,q}(Z)$ has
either (i) a monotonically increasing behavior, (ii) a local minimum, or
(iii) a global minimum (see Fig.~\ref{fig:hkvZ} in the Supplementary
Material). Monotonically increasing behavior corresponds to a second-order 
percolation transition. When $h_{k,q}(Z)$ has a global minima, percolation transition is an abrupt (first-order) transition. The presence of local minima
indicates that the percolation transition is a two-stage transition in
which the giant component undergoes an abrupt (first-order) jump
followed by a continuous transition as the occupation probability $p_0$
is decreased [see the case of $q= 0.765$ in
  Fig.~\ref{fig:simvthr}(c)]. Using this analysis, we plot the complete
phase diagram of \kcore percolation transition for \er networks in
Fig.~\ref{fig:phaseplot}.

The boundaries of the phase diagram (Fig. \ref{fig:phaseplot}), $q = 0$ and $k=1$ lines correspond to the cases of \kcore percolation in single network and regular percolation in interdependent networks, respectively. We describe the complex nature of the
combined \kcore percolation and interdependent network model at
intermediate couplings $0 < q <1$, and contrast it with the known results at the
boundaries. Parshani \textit {et al.}~\cite{Parshani_prl} demonstrated
that regular percolation in coupled networks changes from a second-order
to first-order when it passes through a tricritical point at the
critical coupling $q_{\rm tri,1}$. The tricritical nature is preserved
in \kcore percolation as the average local threshold $k$ is increased,
but the tricritical coupling $q_{{\rm tri},k}$ increases with $k$, as can be
seen in Fig.~\ref{fig:phaseplot}. The dependence of $q_{{\rm tri},k}$ on
the average degree $z_1$ is
\begin{equation} \label{eqn:qtrick_gen}
q_{{\rm tri},k} = 1 + X_{k-1,0} -\sqrt{(1+X_{k-1,0})^2-1},
\end{equation}
where $X_{k-1,0}$ is the numerical solution for $X$ in self-consistent
Eq.~(\ref{eqn:XZ}) when $Z = 0$.

A first-order transition indicates network instability. Because
instability increases with an increase in both the coupling $q$ and the
average local threshold $k$---more nodes are removed during \kcore
percolation at higher local thresholds---we expect the \kcore
percolation transition to become first-order at lower couplings when the
average local threshold is higher. Counterintuitively, Figure~\ref{fig:phaseplot} shows that
the tricritical coupling $q_{{\rm tri},k}$ increases with $k$. To test
this further, we analyse Eq.~(\ref{eqn:qtrick_gen}). A perturbative
expansion shows that $q_{{\rm tri},k}$ indeed increases with
$k$, around $k = 1$, as
\begin{equation} \label{eqn:qtrick}
q_{{\rm tri},k} = q_{\rm tri,1} +
\frac{(k-1)e^{-1}+(k-1)^2e^{-2}}{z_1}\left( \frac{z_1 + 1}{\sqrt{2z_1 +
    1}} - 1\right), 
\end{equation}
where the tricritical coupling $q_{\rm tri,1}$ (consistent with results
found in Ref.~\cite{Jianxi_Nonpre}) is given by
\begin{equation} \label{eqn:qtric1}
q_{\rm tri,1} = 1 +\frac{1}{z_1}-\sqrt{(1+\frac{1}{z_1})^2-1}.
\end{equation}
We compare the perturbative solution of Eq.~(\ref{eqn:qtrick}) with the
numerical solution of Eq.~(\ref{eqn:qtrick_gen}) and the simulation
results in Fig.~\ref{fig:qtrik_comp} (see Supplementary Material).

Above an average local threshold $k \lesssim 2$, the tricritical nature
ceases to exist. Instead, as the coupling $q$ is increased, the \kcore
percolation transition goes through a two-stage transition as it changes
from second-order to first-order. Figure~\ref{fig:simvthr}(c) shows that
this two-stage transition has characteristics of both first- and
second-order transitions.  The critical couplings that separate the
two-stage transition from the first-order and second-order transition
regions are $q_{c,1}$ and $q_{c,2}$, respectively. At the critical line
$q_{c,2}(k)$, the function $h_{k,q}(Z)$ develops an inflection point at
$Z > 0$ that signals the development of a local minimum for $q >
q_{c,2}$ (see Fig.~\ref{fig:hkvZ}(b) in the Supplementary Material). The
condition for $q_{c,2}$ at a fixed $k$ is
\begin{equation} \label{eqn:qc2_cond}
  h^{\prime}_{k,q_{c,2}}(Z_0) = 0   \quad \&  \quad  
h^{\prime\prime}_{k,q_{c,2}}(Z_0) = 0,
\end{equation}
where the derivatives are taken with respect to $Z$ and the inflection
point $Z_0$ must be determined using the relationship in
Eq.~(\ref{eqn:qc2_cond}).  For couplings $q \le q_{c,1}$, the global
minimum of $h_{k,q}(Z)$ occurs at $Z=0$. For $q > q_{c,1}$, the global
minimum shifts to $Z_0 >0$. At the critical line $q_{c,1}(k)$, the
function has global minima at both $Z=0$ and $Z_0>0$ (see
Fig.~\ref{fig:hkvZ}(b) in the Supplementary Material) and this yields
the conditions for the critical coupling $q_{c,1}$,
\begin{equation}\label{eqn:qc1}
 h_{k,q_{c,1}}^\prime(Z_0) = 0   \quad \&  \quad  h_{k,q_{c,1}}(Z_0) =
 h_{k,q_{c,1}}(Z = 0). 
\end{equation}	

In single networks, the k-core percolation transition reaches a
tricritical point when the average local threshold is increased from $2$
to $3$ at $k_c = 2.5$ \cite{tricrit}. Figure~\ref{fig:phaseplot} shows
that this tricritical point is preserved when the coupling between the
networks is increased up to a critical coupling $q_{c,2.5}$ and forms a
second tricritical line. The point $q_{c,2.5}$ (point ``X'') is a triple point surrounded by three phases. This critical coupling  depends
on the average degree $z_1$,
\begin{equation}\label{eqn:qc2.5ER}
q_{c,2.5} = 1 +\frac{3}{2z_1}-\sqrt{(1+\frac{3}{2z_1})^2-1}.
\end{equation}

The critical lines $q_{c,1}(k)$ and $q_{c,2}(k)$ can be calculated perturbatively around the point $q_{c,2.5}$. Using the expansion of $h_{k,q}(Z)$ around $Z = 0$ with the conditions in Eq.(\ref{eqn:qc2_cond}) and Eq.(\ref{eqn:qc1}), we get a general equation

\begin{equation} \label{eqn:qc1qc2_pert}
a_m(1-q)^4
 + b_mq(1-q)^2 + c_mq^2 = 0,
\end{equation}
where $a_m = \frac{z_1^2}{36}(12(3-2m)\delta^2 + 6(m-2)\delta + 1)$, $b_m =\frac{z_1}{6}(12(1-m)\delta^2+(4-2m)\delta-1), c_m = \delta^2 + \delta + 1/4$ with $\delta = 2.5 - k$. Solving Eq.(\ref{eqn:qc1qc2_pert}) with $m = 3$ and $m = 4$ gives $q_{c,2}$ and $q_{c,1}$, respectively. The numerical solution of Eq. (\ref{eqn:qc1qc2_pert}) are plotted in the Supplementary material (see Fig. \ref{fig:qc1qc2pert}).

Finally, for the average local threshold $2.5 < k \le 3$, \kcore percolation
transition remains first-order even when the coupling between the
networks is increased.

The critical percolation thresholds and critical
exponents for all three transitions discussed above can be calculated
from the function $h_{k,q}(Z)$. At the second-order transition and the
continuous part of the two-stage transition ($q < q_{c,1}$, the gray
regions in Fig.~\ref{fig:phaseplot}), the critical behavior of the giant
component takes the form $\phi_{\infty} \sim (p - p_{c,2})^{\beta_2}$,
where $p_{c,2}=h_{k,q}(Z=0)$. The analytical expressions for $p_{c,2}$
are
\begin{equation}\label{eqn:pc2}
p_{c,2} = \left\{%
\begin{array}{lcrcl}
\frac{1}{z_1(1-q)}, &1 \le k \le 2&\\        
 \frac{1}{z_1(1-(k-2))(1-q)},  &2 \le k \le 2.5&\\
\end{array}
\right.
\end{equation}

\begin{figure}
\centering
\includegraphics[width=\columnwidth]{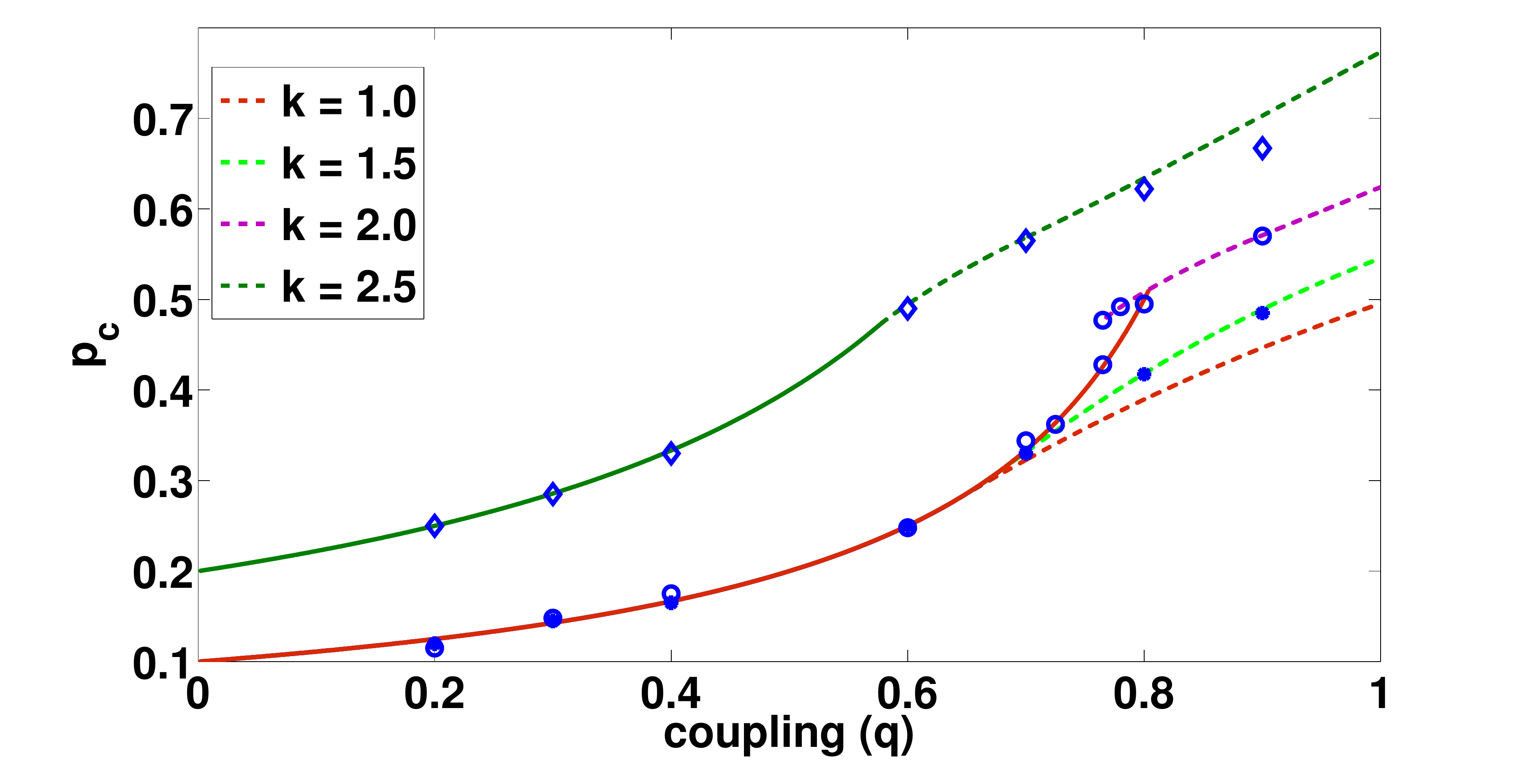}
\hspace{.0in}
  \includegraphics[width=\columnwidth]{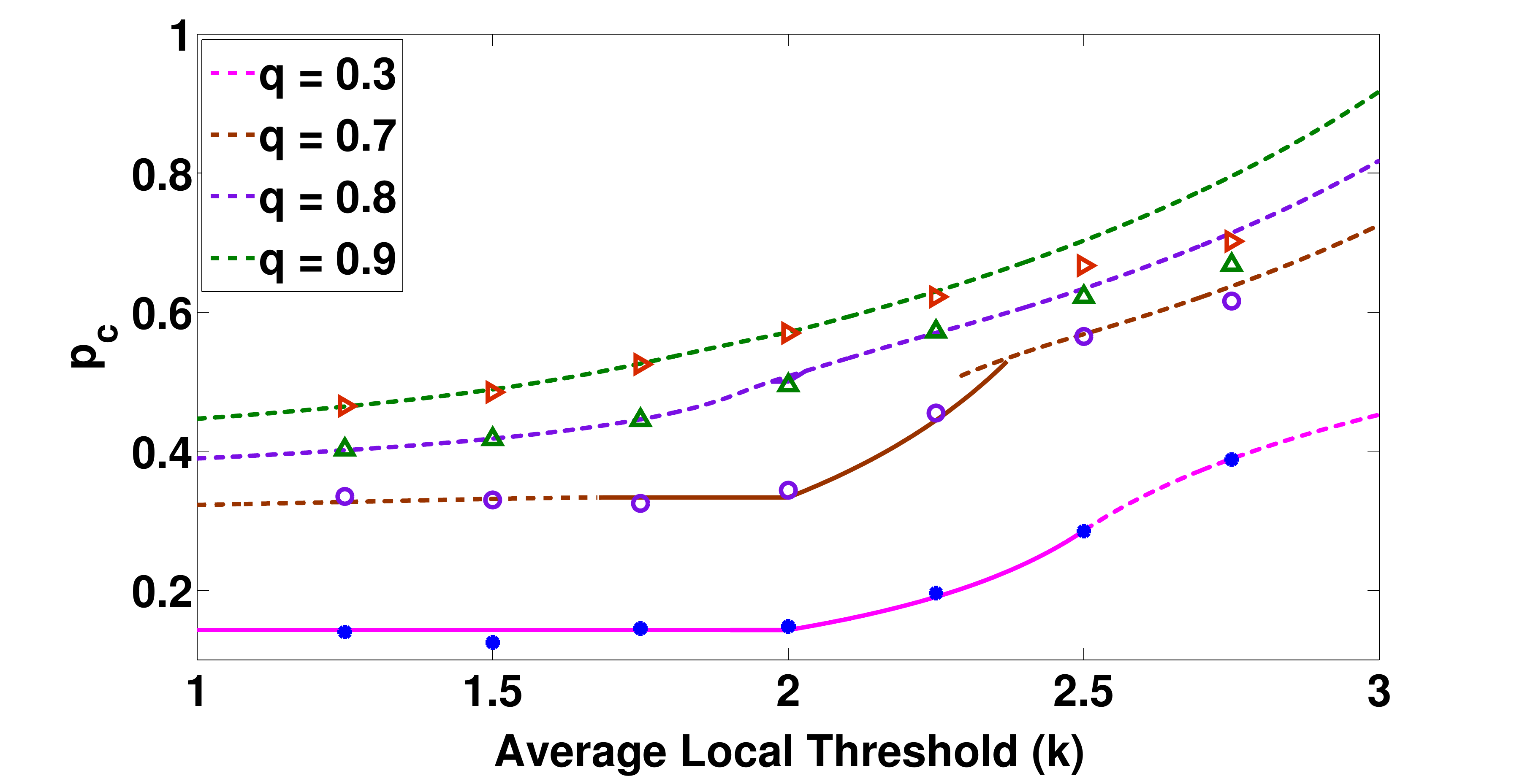}
\caption{Percolation threshold $p_c$ as a function of (a) the coupling
  $q$ for fixed average local threshold $k = 1.0,1.5,2.0,2.5$
  representing horizontal lines in Fig. \ref{fig:phaseplot} (b) the
  average local threshold $k$ for several fixed coupling $q =
  0.3,0.7,0.8,0.9$ representing vertical lines in
  Fig. \ref{fig:phaseplot}. Dashed and continuous lines indicate that
  the percolation threshold is at abrupt (first-order) jump and
  continuous transition respectively. Simulation results (shown as symbols) are obtained for a system with $10^6$ nodes in each network.} \label{fig:pcvL_comb}
\end{figure}

We find the exponent $\beta_2$ by using the Taylor series expansion of
the function $h_{k,q}(Z)$ around $Z=0$. The exponent depends on
coupling, indicating that coupling changes the universality classes of
these \kcore percolation transitions. The exponents found at different points of the
phase diagram are
\begin{equation}\label{eqn:beta_2}
\beta_2 = \left\{
\begin{array}{lcrcl}
1, &1 \le k < 2, q < q_{{\rm tri},k}&\\
1/2, &1 \le k < 2, q = q_{{\rm tri},k}&\\
 2, &2 \le k < 2.5, q \le q_{c,1}&\\
1, &k = 2.5, q < q_{c,2.5}&\\
2/3, &k = 2.5, q = q_{c,2.5}.&
\end{array}
\right.
\end{equation}
At the first-order transition and the abrupt jump of the two-stage
transition, the critical behavior of the giant component takes the form
$\phi_{\infty} - \phi_{\infty,0} \sim (p - p_{c,1})^{\beta_1}$, where $p_{c,1}= h_{k,q}(Z_0)$.
$Z_0$ is the minimum of the function $h_{k,q}(Z)$ found using the condition $ h_{k,q}^\prime(Z_0) = 0$. Both $p_{c,1}$
and $p_{c,2}$ are calculated numerically and are in good agreement with
the simulations shown in Fig.~\ref{fig:pcvL_comb}. We calculate the
critical exponent $\beta_1$ using a Taylor series expansion of the
function $h_{k,q}(Z)$ around the minimum $Z_0$ and find that it is
dependent only on coupling $q$
\begin{equation}\label{eqn:beta_1}
\beta_1 = \left\{%
\begin{array}{lcrcl}
1/3,  &q = q_{c,2}& \\
1/2,  &q > q_{c,2}&.
\end{array}
\right.
\end{equation}

The richness of the phase diagram is striking when the change in \kcore percolation transition is considered as threshold $k$ is increased at fixed $q$. At certain fixed intermediate couplings, the \kcore percolation transition changes from first-order $\rightarrow$ second-order $\rightarrow$ two-stage $\rightarrow$ first-order as the \kcore threshold is increased (See vertical arrow in Fig. \ref{fig:phaseplot}). Additionally, note that the result for fully interdependent networks $q=1$ is
consistent with the result for the \kcore percolation transition in multiplex
networks in that they are both first-order for any average threshold $k$
\cite{multiplexkcore}.

In conclusion, we have developed and analysed a general model that includes two
realistic mechanisms: \kcore percolation and interdependency between
networks with any degree of coupling. We have verified our analytical
solutions through extensive simulations. We have demonstrated the richness of combined effects through the
complete phase diagram for \kcore percolation transition in two
interdependent \er networks. The coupling between networks dramatically
changes the critical behavior of \kcore percolation found in single
networks, and also yields new critical exponents that are calculated
analytically. At fixed \kcore threshold, the \kcore percolation transition changes from
second-order to first-order as the coupling is increased, either passing
through a tricritical point or two-stage transition depending on the
average local threshold. We calculated the tricritical couplings and phase
boundaries of the two-stage transition shared with
second and first-order transition regions. Counterintuitively, we find the tricritical coupling to increase with the \kcore threshold. The richness of this generalized model is further emphasized with the \kcore percolation transition, for certain fixed couplings, changing from first-order $\rightarrow$ second-order $\rightarrow$ two-stage $\rightarrow$ first-order as the \kcore threshold is increased, in contrast to second-order $\rightarrow$ first-order for single networks. To test the universality of our results,
we also analyzed, both analytically and numerically, the phase diagram
for \kcore percolation in interdependent random regular networks and
found this system to be very similar to that of \er
networks (See Supplementary Material). Studying
these new percolation transitions found in this generalized model  
will enable us to 
understand the importance and the rich effects of coupling between different resources in cascading failures
that occur in real world systems, which will enable us to design more resilient systems.

\begin{acknowledgements}
We thank the financial support of the Office of Naval
Research Grants N00014-09-1-0380, N00014-12-1-0548 and
N62909-14-1-N019; the Defense Threat Reduction Agency Grants
HDTRA-1-10-1-0014 and HDTRA-1-09-1-0035; National Science
Foundation Grant CMMI 1125290 and the U.S.- Israel Binational
Science Foundation- National Science Foundation Grant 2015781;
the Israel Ministry of Science and Technology with the Italy Ministry of
Foreign Affairs; the Next Generation Infrastructure (Bsik); and the
Israel Science Foundation.
\end{acknowledgements}

\widetext
\clearpage
\begin{center}
\textbf{\large Supplemental Material:  \kcore percolation in interdependent networks}
\end{center}

\setcounter{equation}{0}
\setcounter{figure}{0}
\setcounter{table}{0}
\setcounter{page}{1}
\makeatletter
\renewcommand{\theequation}{S\arabic{equation}}
\renewcommand{\thefigure}{S\arabic{figure}}
\renewcommand{\bibnumfmt}[1]{[S#1]}
\renewcommand{\citenumfont}[1]{S#1}

\section{C\lowercase{omparison of} G\lowercase{iant component from theory and simulations for two coupled scale-free networks}}

\begin{figure}[H]
 \begin{center}
\includegraphics[width=0.5\columnwidth]{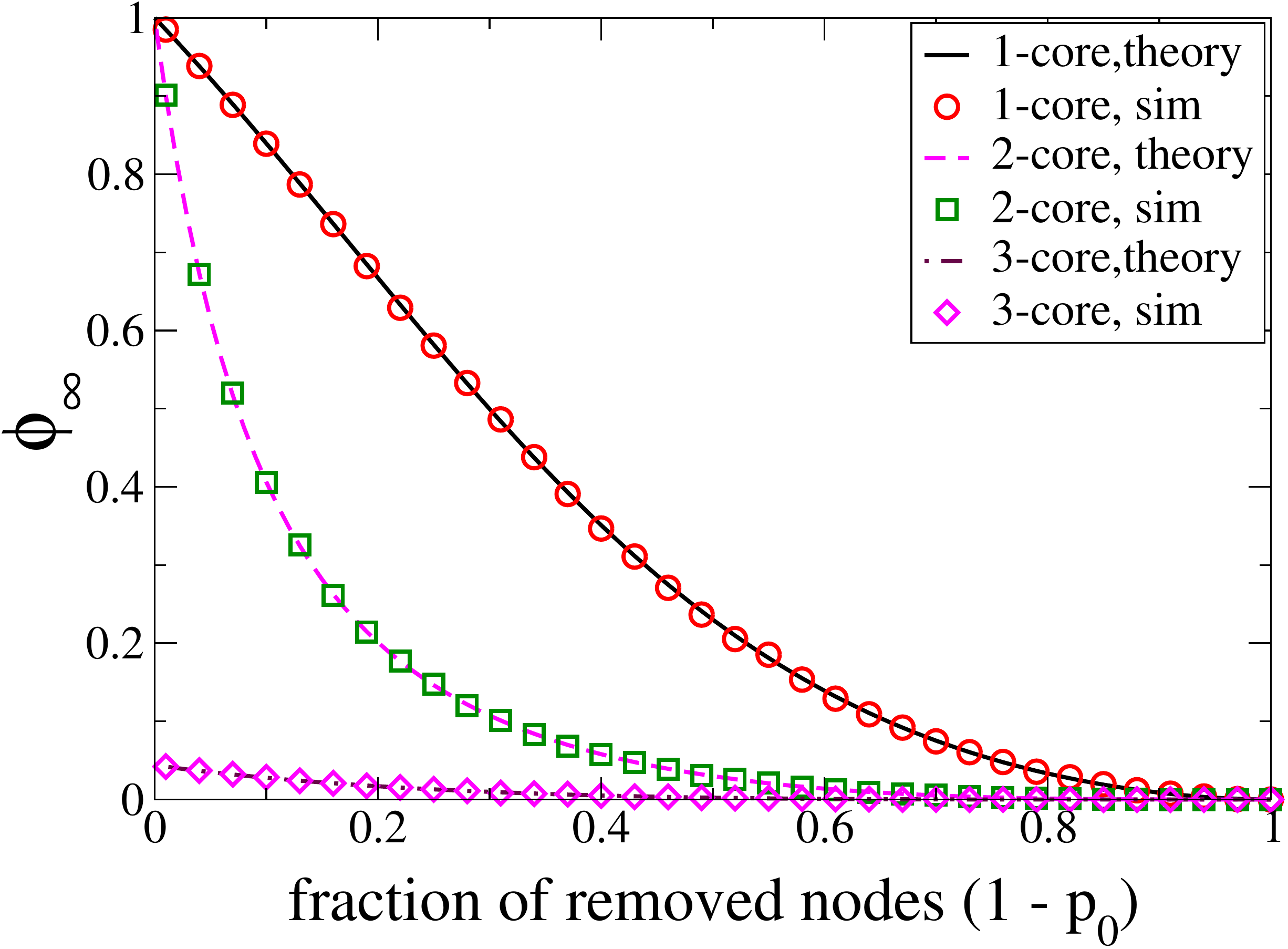}

\caption{ 
Comparison between theory and simulation of \kcore percolation in two interdependent Scale-free networks with exponent $\gamma = 2.5$, with both layers having identical local thresholds $k=1,2,3$. Simulation results were obtained for a system with $N = 10^6$ nodes in each network. The minimum and maximum degree for nodes in each network were set to be $i_{min} = 2$ and $i_{max} = 1000$, respectively.} \label{fig:simvthr_sf}
\end{center}
\end{figure}

\newpage

\section{C\lowercase{omparison of} G\lowercase{iant component from theory and simulations for two coupled} \er \lowercase{networks}}

\begin{figure}[H]
\begin{center}
\mbox{ 
\subfigure{
\includegraphics[width=0.4\columnwidth]{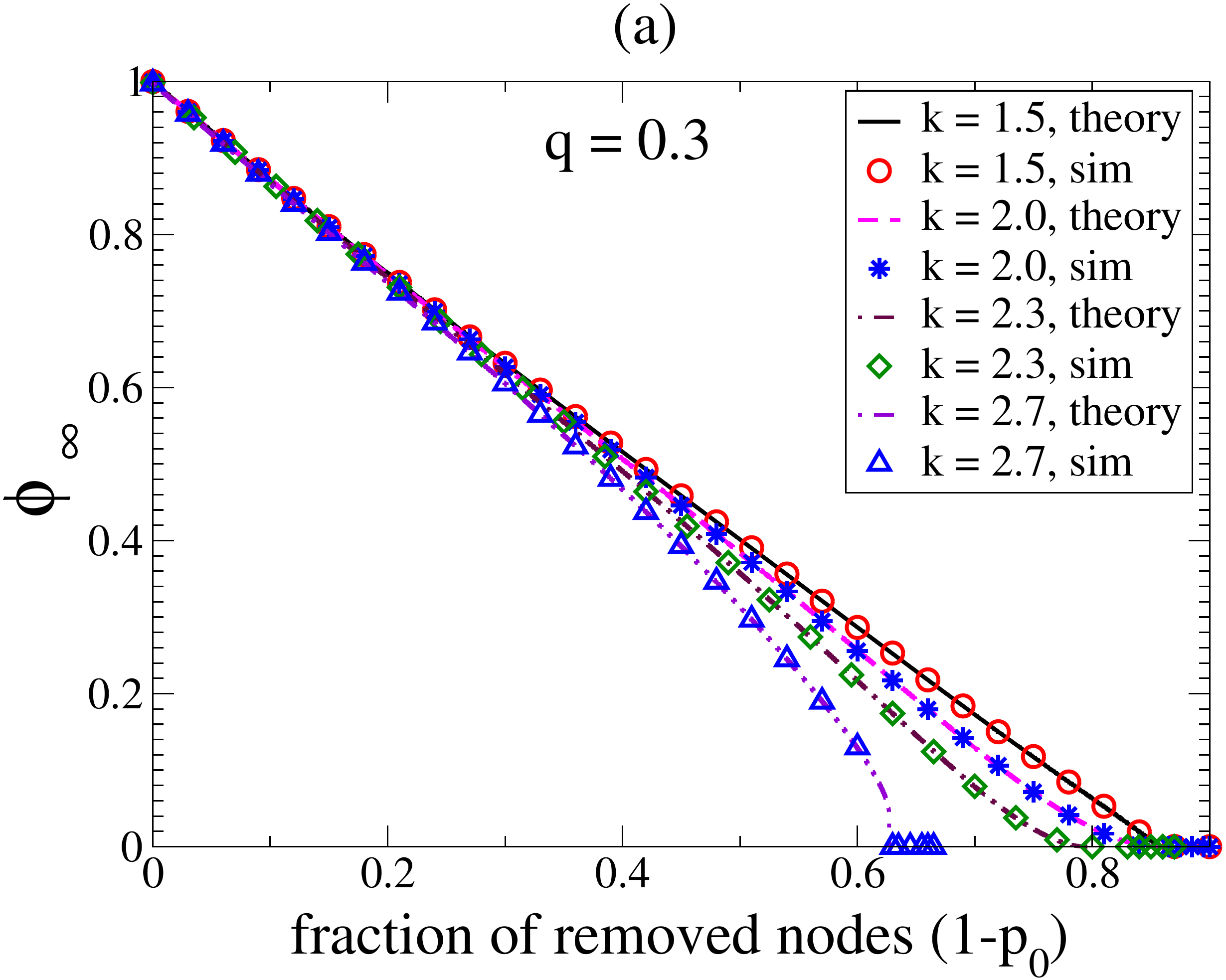}}	

\subfigure
{
\includegraphics[width=0.4\columnwidth]{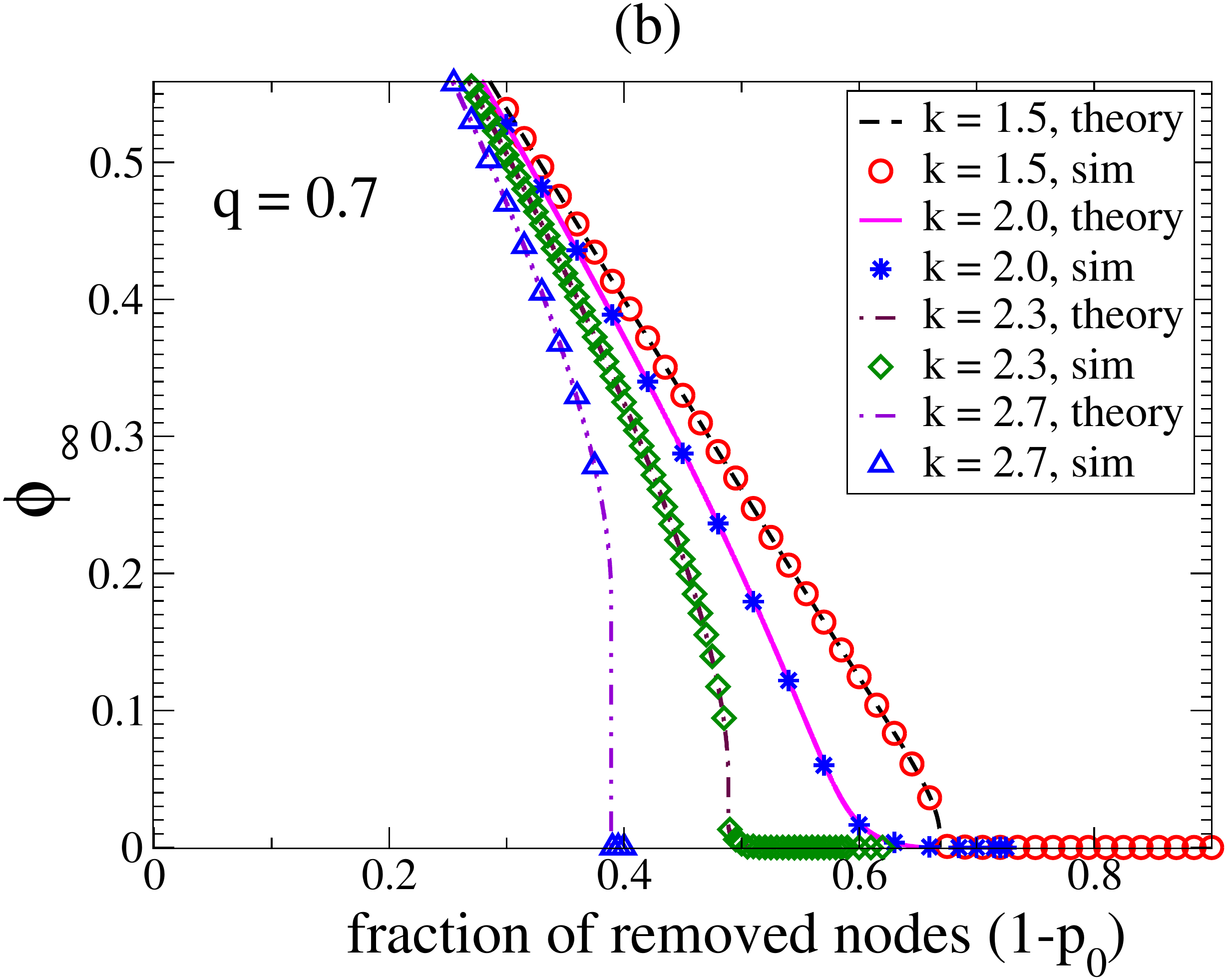}}	
}
\newline
\mbox{
\subfigure{
\includegraphics[width=0.4\columnwidth]{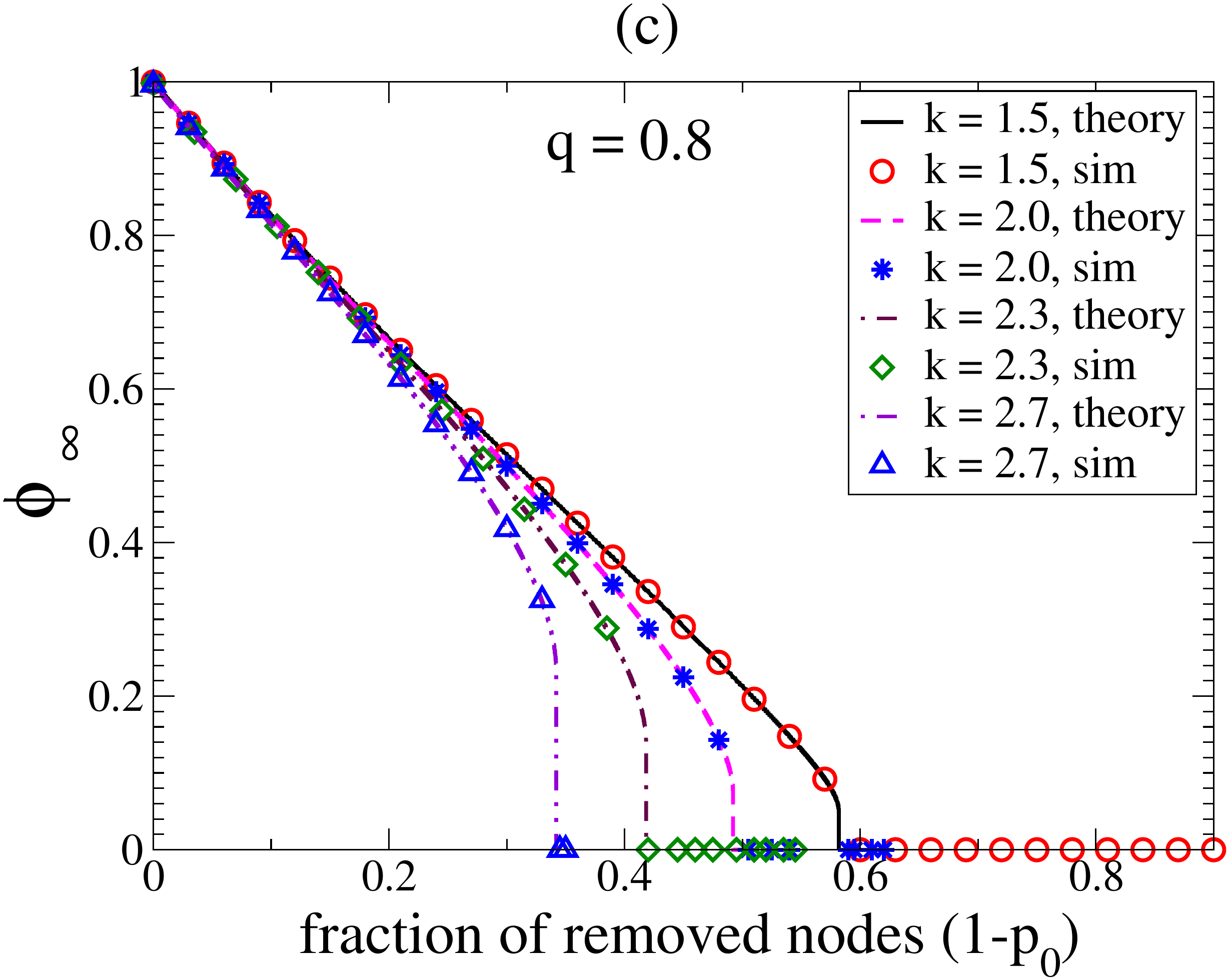}}

\subfigure{
\includegraphics[width=0.4\columnwidth]{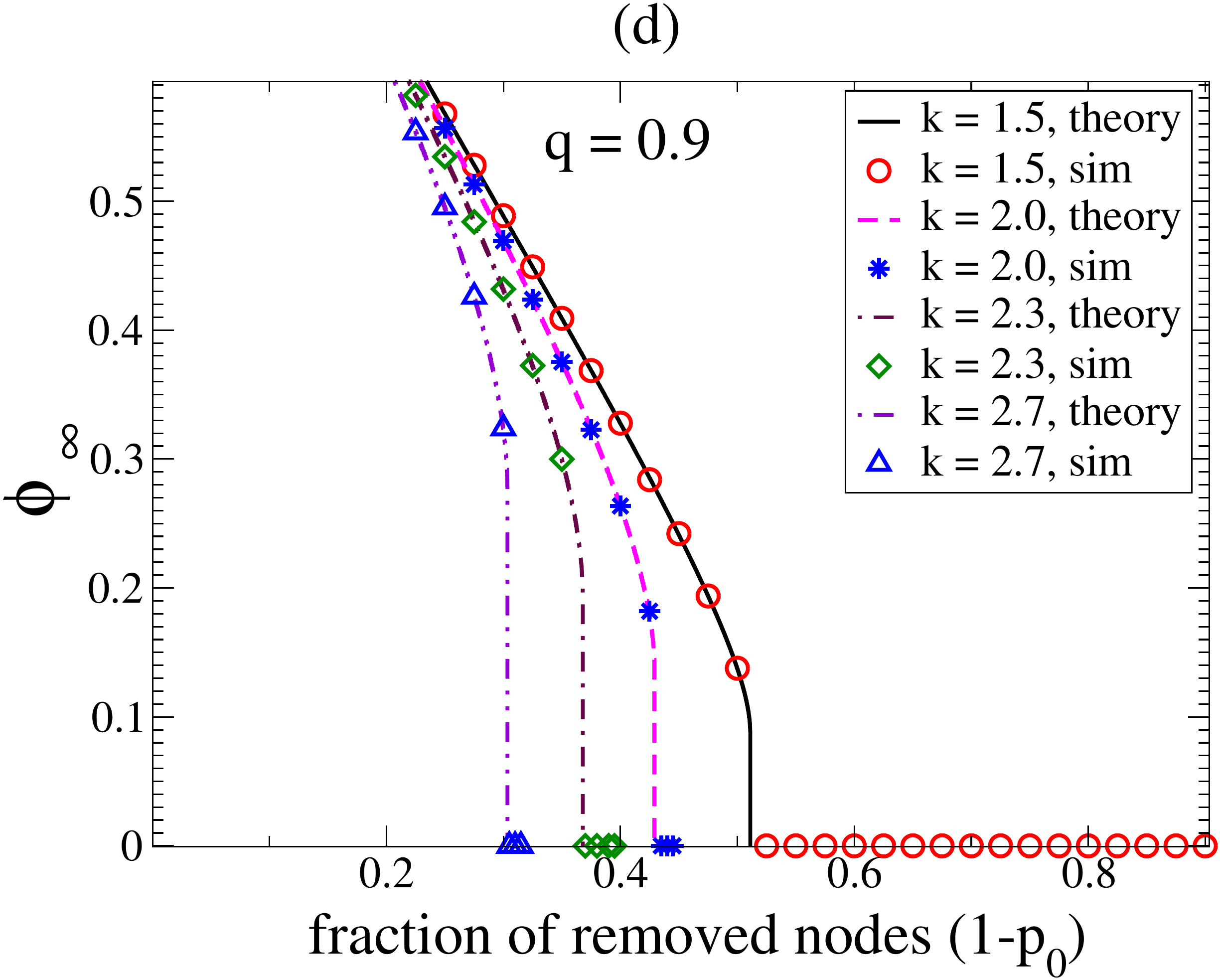}}	

}
\end{center}
\caption{The giant component for two coupled \er networks ($z_1 = 10$), computed numerically and through simulations, as a function of fraction of initially removed nodes $p_0$ at different average local threshold $k$ for couplings a) $q = 0.3$ b) $q = 0.7$ c) $q = 0.8$ and d) $q = 0.9$. For low coupling $q = 0.3$, nature of \kcore percolation is identical to that of single networks. For high couplings $q = 0.8$ and $q= 0.9$, \kcore percolation is first-order indicating the increased instability of the system compared to single networks. For the intermediate coupling $q = 0.7$, \kcore percolation is initially first-order for $k = 1.5$, which then becomes a continuous transition as the average local threshold is increased to $k = 2.0$. The cascades during \kcore percolation are expected to increase as the local threshold of nodes are increased, and therefore, \kcore percolation would be (intuitively) expected to remain as first-order. As the average local threshold is increased further, the increased instability in the system is manifested into \kcore percolation becoming a two-stage transition at $k = 2.3$ and finally into a first-order transition for $k = 2.7$. Simulation results (shown as symbols) are obtained for a system with $10^6$ nodes in each network.} \label{fig:Svpfixedq}
\end{figure}

\newpage
\section{C\lowercase{omparison of behavior of the function $h_{k,q}(Z)$ at tricritical point and two-stage transition}}
\begin{figure}[H]

\begin{center}
\mbox{ 
\subfigure
{
\includegraphics[width=0.9\columnwidth]{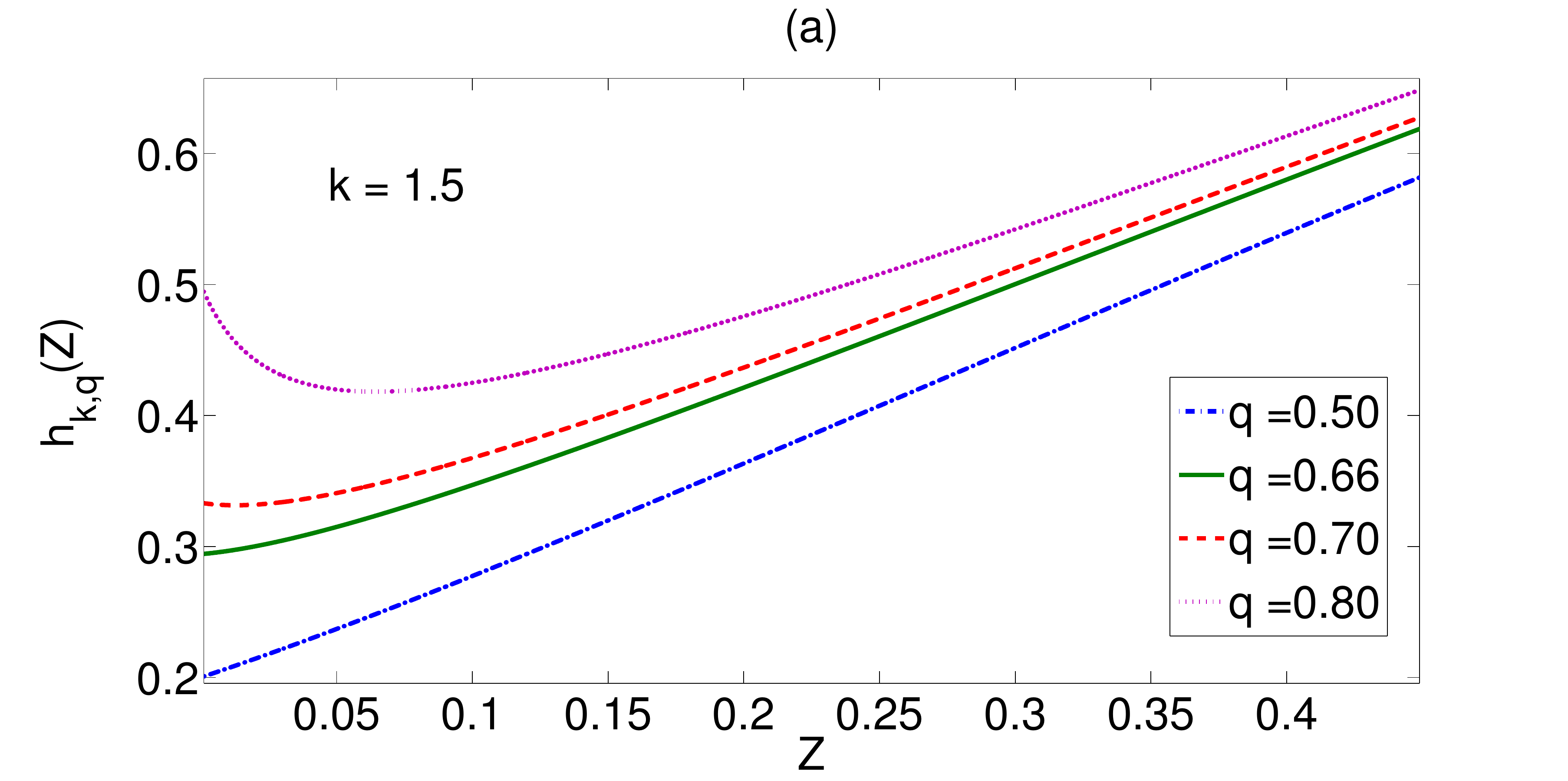}}
}
\newline
\mbox{
\subfigure
{
\includegraphics[width=0.9\columnwidth]{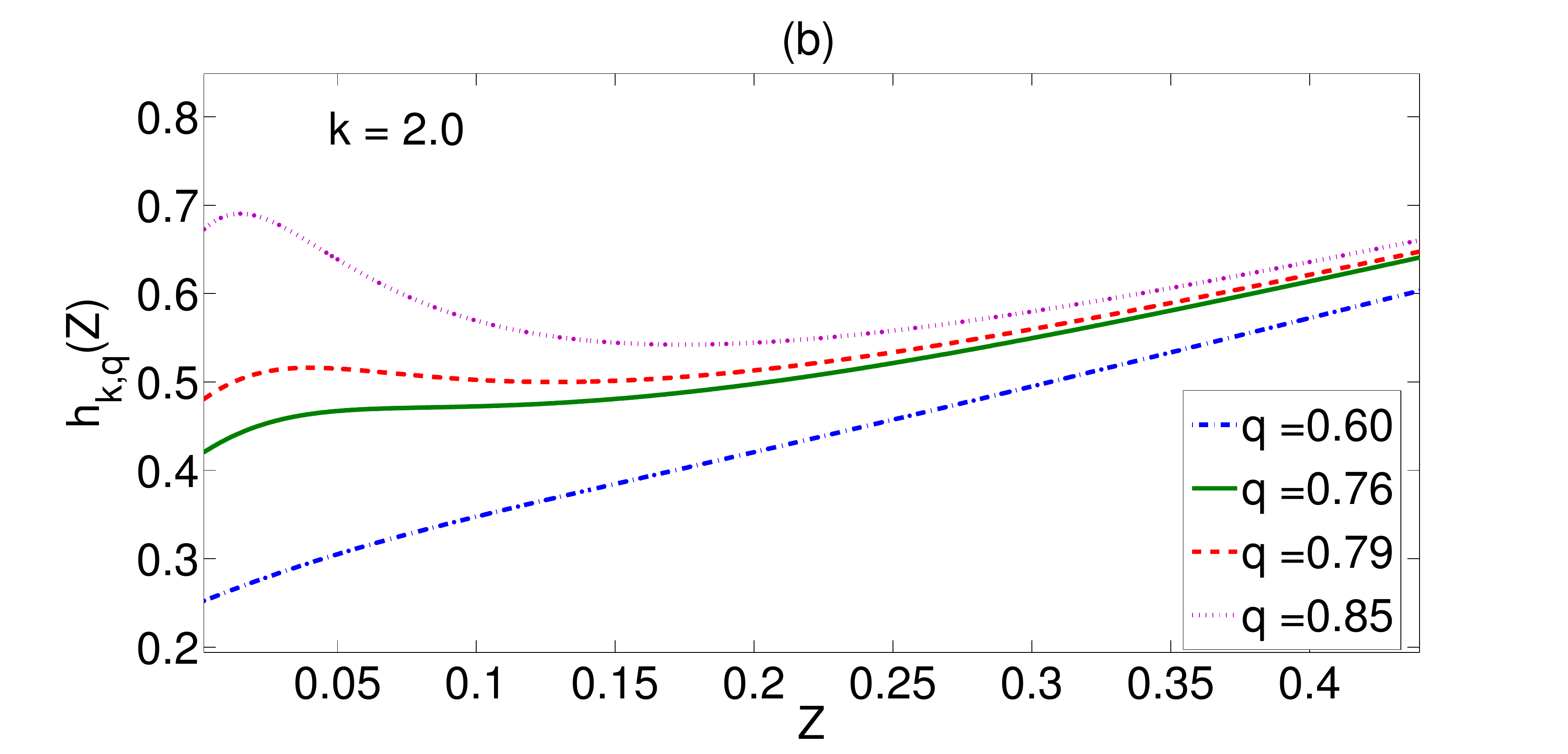}}
}
\end{center}

\caption{Comparison of behavior of the function $h_{k,q}(Z)$ for two coupled \er networks at fixed average local threshold a) $k = 1.5$ and b) $k = 2.0$ at different couplings. As seen in the phase diagram (See Fig.\ref{fig:phaseplot_diffz_ER}), \kcore percolation changes from a second-order at low couplings to a first-order at high couplings $q$ passing through a tricritical point for $k = 1.5$, and through a two-stage transition for $k = 2.0$. In both cases, $h_{k,q}(Z)$ is characterized by monotonically increasing behaviour corresponding to second-order transition and, by the presence of a global minima corresponding to first-order transition. For $k = 1.5$, the inflection point occurs at $Z=0$, which immediately turns into a global minima as the coupling is increased, leading to a tricritical point. For $k = 2.0$, the inflection point occurs at $Z > 0$, which turns into a local minima followed by being a global minima as the coupling is increased, leading to a two-stage transition.}  \label{fig:hkvZ}
\end{figure}

\newpage
\section{P\lowercase{lot of tricritical coupling as a function of average local threshold for $1<k<2$}}
\begin{figure}[H]
\begin{center}
\includegraphics[width=\columnwidth]{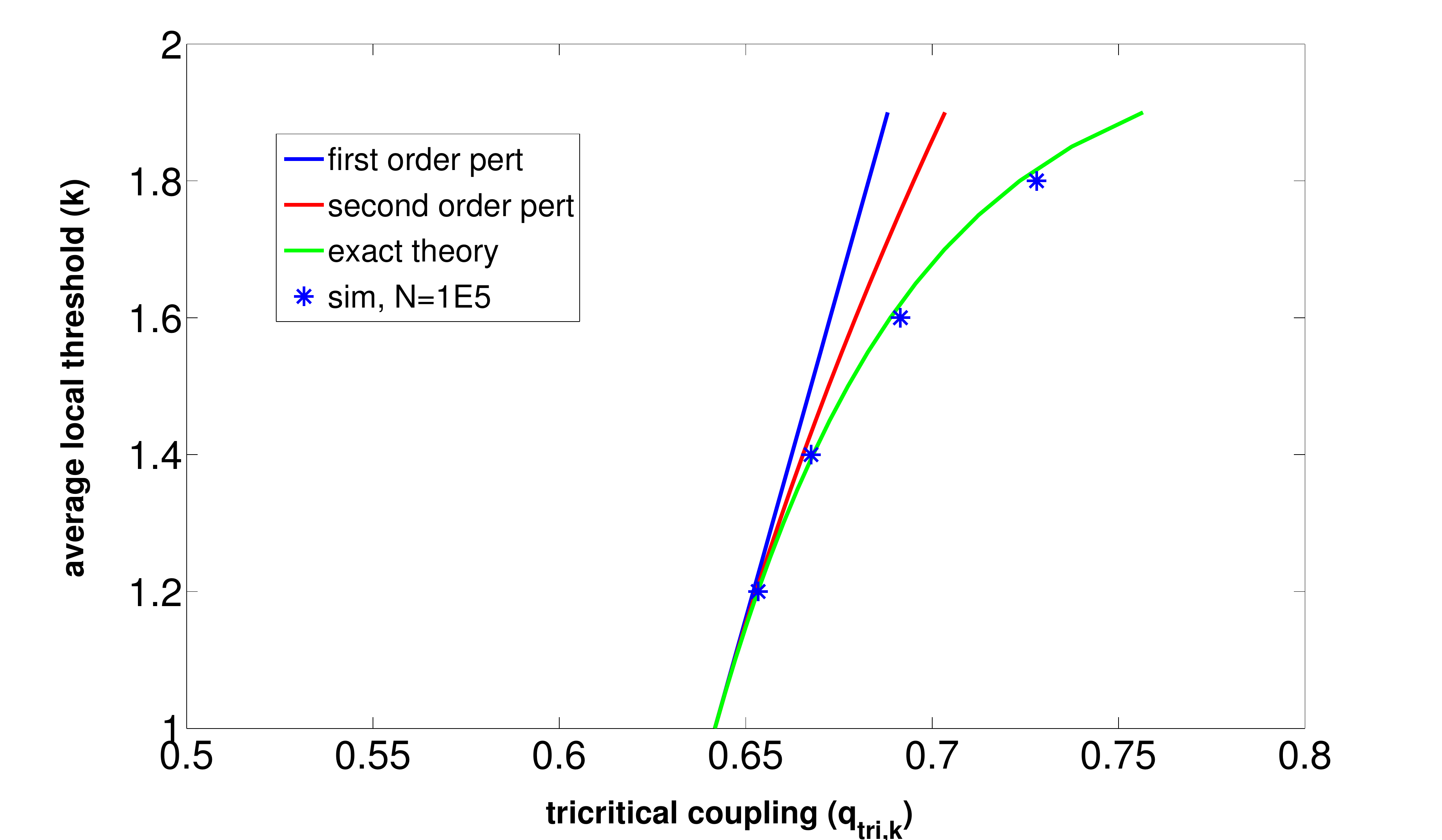}

\caption{Plot of tricritical coupling $q_{{\rm tri},k}$ as a function of average threshold $k$ obtained from the numerical solution of perturbative expansion to first order, second order and exact equation given in the Eqs. (\ref{eqn:qtrick_gen}, \ref{eqn:qtrick}) in the main text. The numerical results are in excellent agreement with the simulation results (shown as symbols) for a network with $10^5$ nodes .}  \label{fig:qtrik_comp}
\end{center}
\end{figure}

\section{P\lowercase{erturbative solution for $q_{c,1}$ and $q_{c,2}$ around the triple point $q_{c,2.5}$}}
\begin{figure}[H]

\begin{center}

\mbox{
\subfigure
{
\includegraphics[width=\columnwidth]{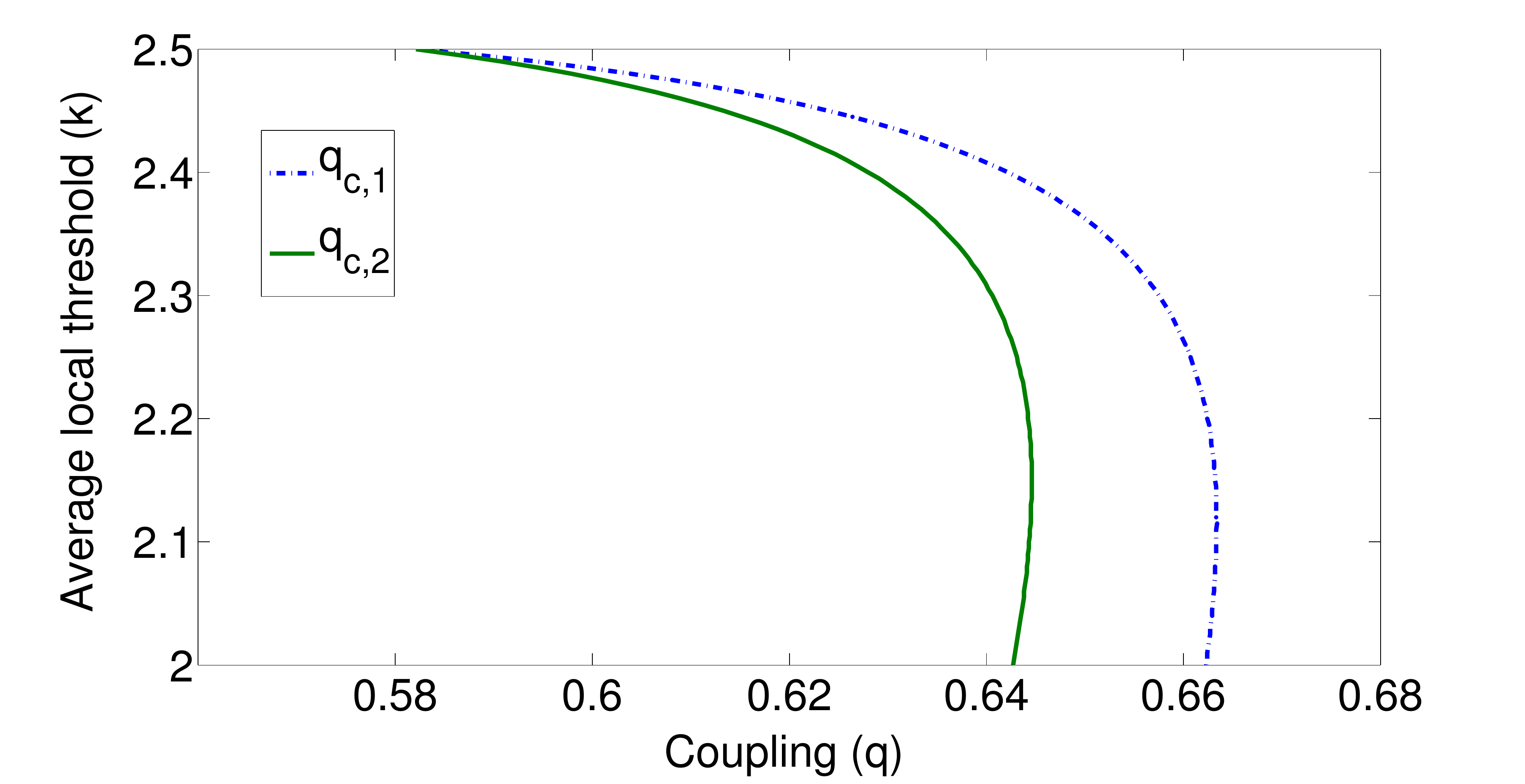}}
}

\end{center}
\caption{Numerical solution of the pertubrative expansion of $q_{c,1}(k)$ and $q_{c,2}(k)$ around the triple point $q_{c2,.5}$ given in Eq. (\ref{eqn:qc1qc2_pert}) in the main text.} \label{fig:qc1qc2pert}

\end{figure}

\newpage
\section{P\lowercase{hase diagram for two coupled} \er \lowercase{networks for different average degree $z1$}}
\begin{figure}[h!]
\begin{center}
\mbox{ 
\subfigure
{
\includegraphics[width=0.5\columnwidth]{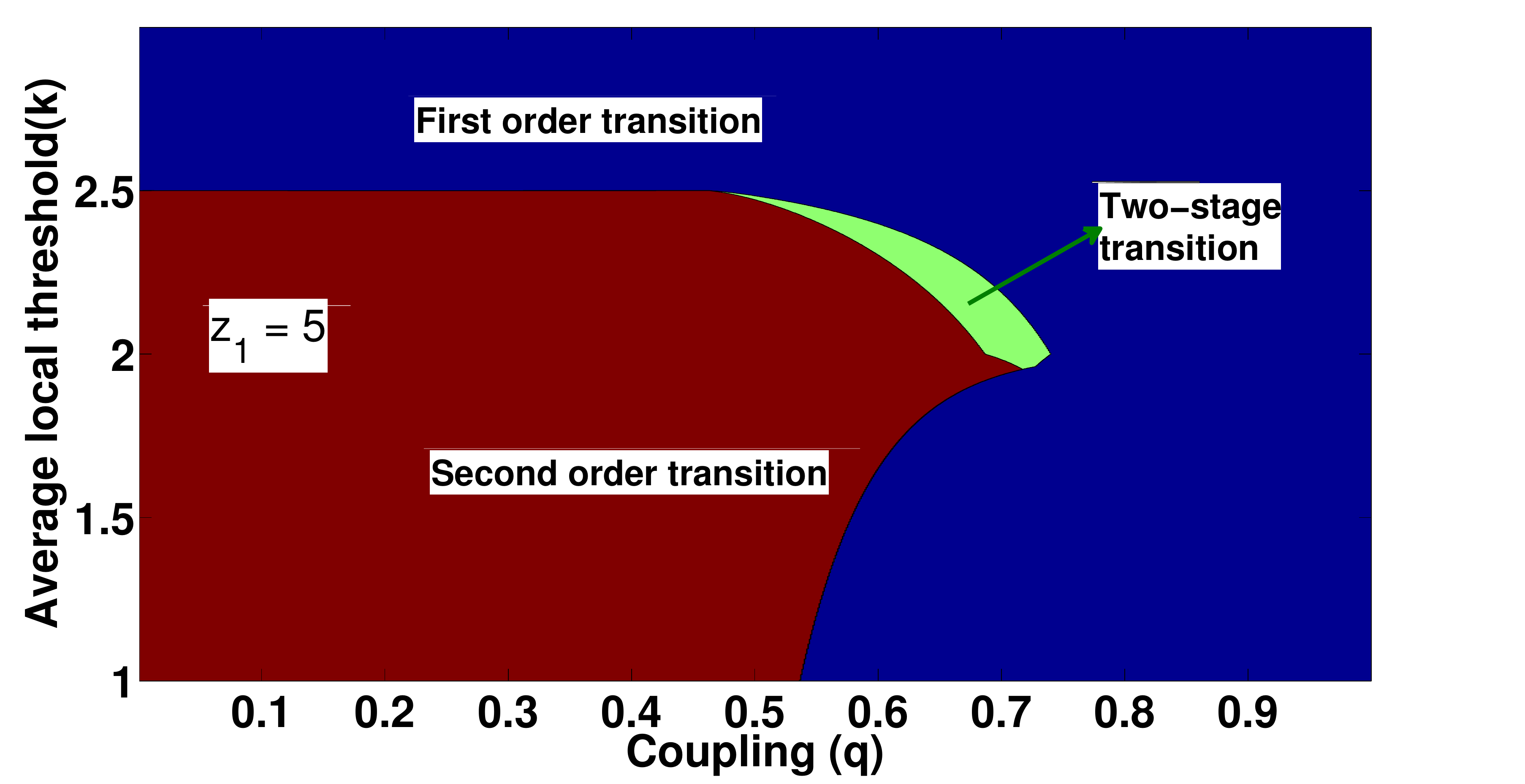}}	

\subfigure
{
\includegraphics[width=0.5\columnwidth]{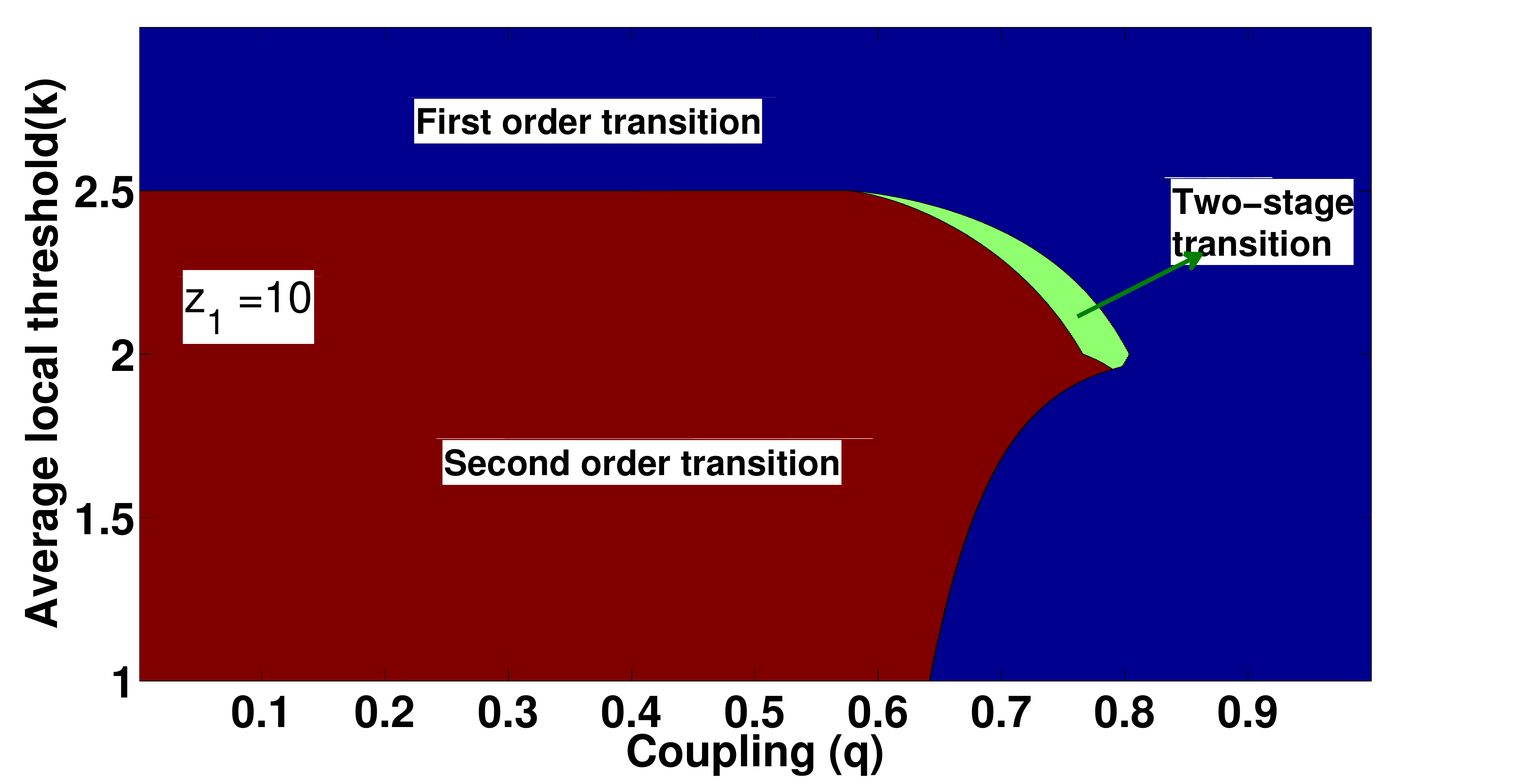}}	
\newline
}
\mbox{
\subfigure
{
\includegraphics[width=0.5\columnwidth]{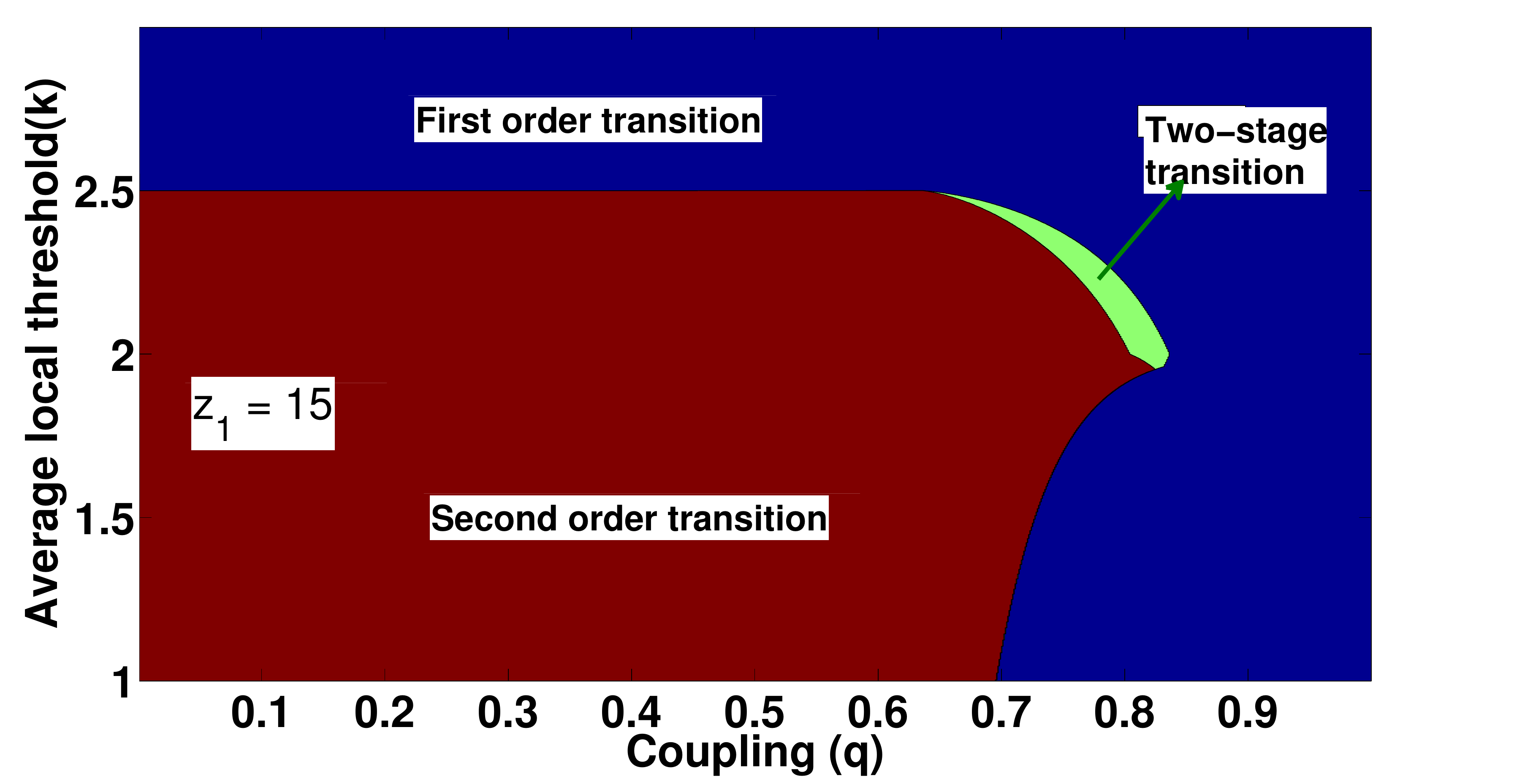}}

\subfigure
{
\includegraphics[width=0.5\columnwidth]{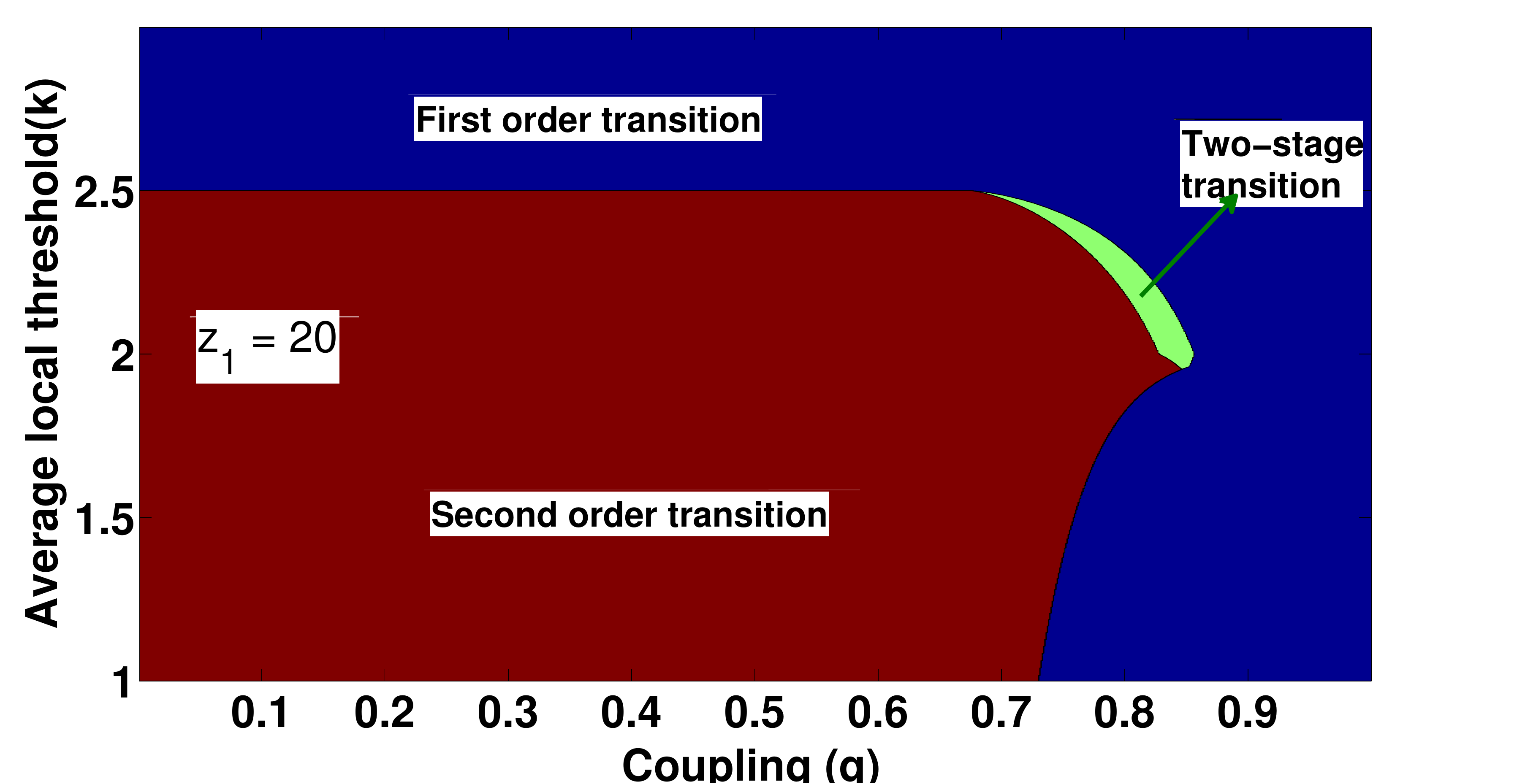}}	

}
\end{center}
\caption{Complete phase diagram for \kcore percolation transition for
  two coupled \er networks with average degree $z_1 = 5, 10, 15, 20$. Both the networks have the
  same average local threshold per site $k=(1-r)k_a+r(k_a+1)$, with fraction $1-r$ of nodes having local threshold $k_a$
  and fraction $r$ of nodes having local threshold $k_a+1$. The transition
  properties depend on the composition of the $k_a$-cores and not
  on average threshold $k$. Width of the two-stage transition region in the phase diagram decreases as the average degree $z_1$ is increased.} \label{fig:phaseplot_diffz_ER}
\end{figure}

\newpage
\section{R\lowercase{andom regular network:} C\lowercase{omplete phase diagram}}

\begin{figure}[H]
 \begin{center}  
  \includegraphics[scale=0.25]{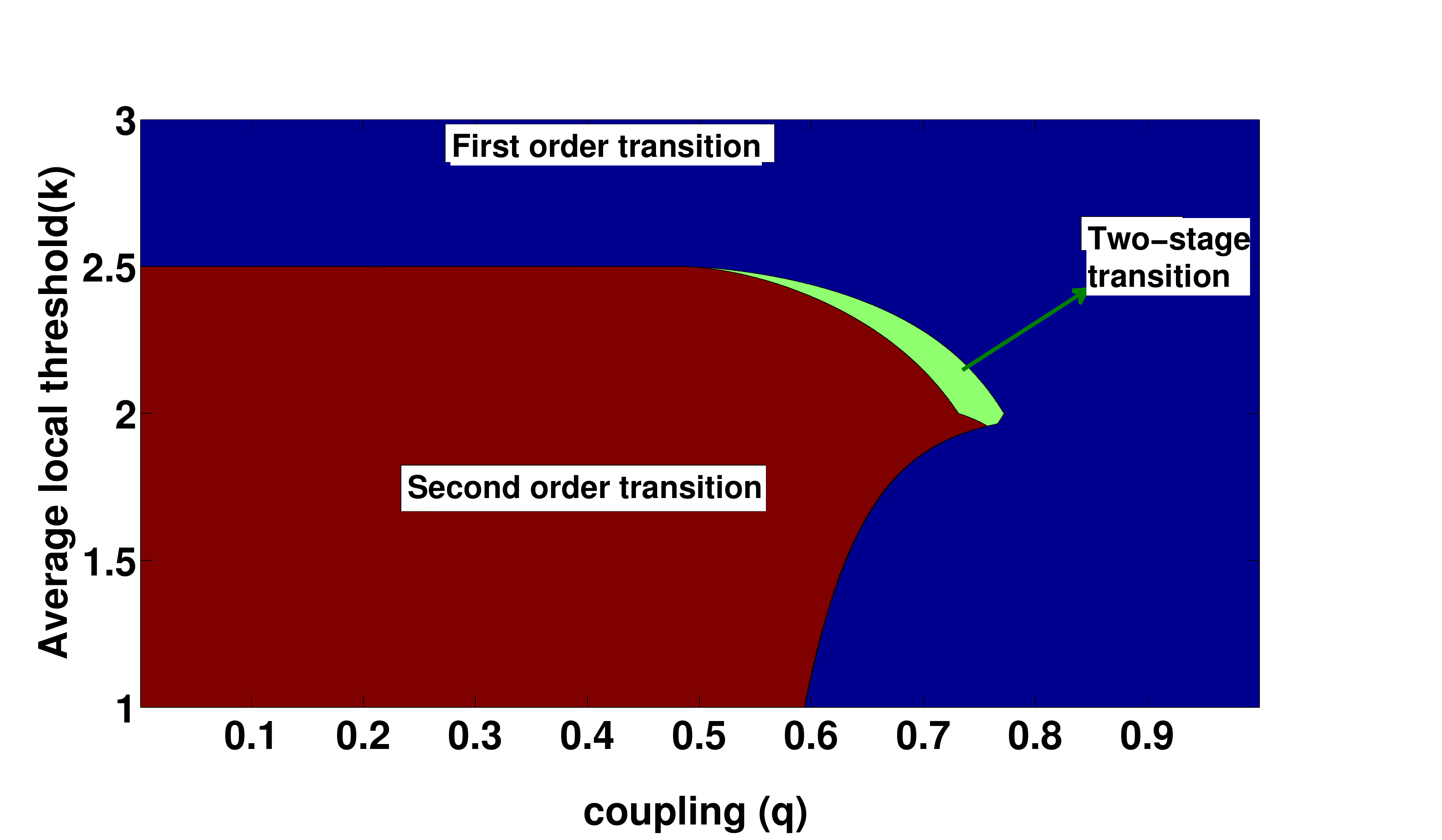} 
\caption{Complete phase diagram for \kcore percolation transition for
  two coupled Random Regular networks with coupling $q$. Both the networks have the
  same local \kcore threshold distribution. A fraction $r$ of randomly chosen nodes have local threshold $k_a+1$ and remaining nodes have $k_a$, resulting in average local threshold $k=(1-r)k_a+r(k_a+1)$. The phase diagram has similar features that were seen in two coupled \er networks(Fig. \ref{fig:phaseplot_diffz_ER}). The critical exponents for all the regions in the phase diagram are identical to that of \er networks as reported in the main text. The expressions for critical percolation thresholds for continuous transition part of both second-order and two-stage transitions are given in Eq. (\ref{eqn:pc2RR}).} \label{fig:phaseplotRR}
\end{center} 
\end{figure}

Consider two coupled Random Regular networks with identical degrees $z_1$. The function $f_{k_a,r}$ is given by $f_{1,r}(X,Z) = 1 - (1-Z)^{z_1-1}                                                                                                 $, $f_{1,r}(X,X) = 1 - r(1-X)^{z_1-1} $,  and since $X = Z$ for $k_a \ge 2$, $f_{2,r}(Z,Z) = 1 - (1-Z)^{z_1-1} - rZ(z_1-1)(1-Z)^{z_1-2}$. The functions $M_{k_a,r}$ are given by $M_{1,r}(X,Z) = 1-(1-Z)^{z_1}-rz_1Z(1-X)^{z_1-1}$  and $M_{2,r}(Z) = 1-(1-Z)^{z_1}-z_1Z(1-Z)^{z_1 -1} - r\frac{z_1(z_1-1)}{2}Z^2(1-Z)^{z_1-2} $. Based on the behavior of $h_{k,q}(Z)$, the complete phase diagram for the percolation transition is plotted in Fig. \ref{fig:phaseplotRR}. The features of the phase diagram are the same as those of coupled \er networks, including identical critical exponents. The critical percolation thresholds are different and, for second-order and continuous part of the two-stage transitions for Random Regular networks is given by,

\begin{equation}\label{eqn:pc2RR}
p_{c,2} = \left\{%
\begin{array}{lcrcl}
\frac{1}{(z_1-1)(1-q)} \color{black}, &1 \le k \le 2&\\ 
 \frac{1}{(z_1-1)(1-(k-2))(1-q)},  &2 \le k \le 2.5&\\
\end{array}
\right.
\end{equation}

The tricritical coupling for regular percolation in interdependent Random Regular networks depends on its degree $z_1$ as given in Eq. (\ref{eqn:qtric1RR}).

\begin{equation} \label{eqn:qtric1RR}
q_{c,1} =1 +\frac{z_1}{(z_1-1)(z_1-2)} 
   -\sqrt{(1+\frac{z_1}{(z_1-1)(z_1-2)})^2-1} .
\end{equation}

The tricritical point found for average local threshold $k = 2.5$ in single RR network is preserved in coupled networks as well.  The tricritical nature persists only up to a critical coupling $q_{c,2.5}$ and its dependence on the degree $z_1$ is given by Eq. (\ref{eqn:qc2.5RR}).
\begin{equation} \label{eqn:qc2.5RR}
q_{c,2.5} = 1 +\frac{3z_1}{2(z_1-2)(z_1-3)}
   -\sqrt{(1+\frac{3z_1}{2(z_1-2)(z_1-3)})^2-1}.
\end{equation}

\end{document}